\documentclass[a4paper,11pt]{article}
\pdfoutput=1 

\usepackage{jheppub}

\usepackage[mmddyyyy,hhmmss]{datetime}

\usepackage{amsmath,graphicx,amssymb,subfigure,bbm,comment}
\usepackage[all]{xy}

\def\<{\langle}
\def\>{\rangle}

\allowdisplaybreaks

\newcommand{\tr}[1]{\text{Tr}\big[#1\big]}

\title{Entanglement Entropy of the ${\mathcal N}=4$ SYM spin chain}

\author{George Georgiou$^a$,}
\author{Dimitrios Zoakos$^b$}

\affiliation{$^a$ Institute of Nuclear and Particle Physics, National Center for Scientific Research Demokritos, 15310 Athens, Greece}
\affiliation{$^b$ Universitat Internacional de Catalunya,
Immaculada 22, E-08017 Barcelona, Spain}

\emailAdd{georgiou@inp.demokritos.gr}
\emailAdd{zoakos@gmail.com}

\abstract{We present a detailed study of the Entanglement Entropy (EE) of excited states in 
all closed rank one subsectors of  ${\cal N}=4$ SYM, namely $SU(2)$, $SU(1|1)$ and $SL(2)$.
Exploiting the techniques of the Coordinate and the Algebraic Bethe Ansatz we obtain the EE for 
spin chains with up to seven magnons, at leading order in the coupling expansion 
but exact in the length of the spin chain and of the part of it that we cut.  
Focusing on the superconformal primary operator with two magnons in the BMN limit, 
we derive analytic and exact, in the coupling $\lambda'$, expressions for the Renyi and the EE.
The interpolating functions for the Renyi and the EE monotonically increase  as the coupling increases from 
 the weak coupling $\lambda' \rightarrow 0$ regime
to the strong coupling $\lambda' \rightarrow\infty$ regime. This results to a violation of a certain bound for the EE that is present 
at weak coupling and confirms the physical intuition that entanglement increases when the coupling increases.}

\begin{document}
\maketitle
\flushbottom

\section{Introduction}
\label{Intro}

 One of the most intriguing  features of a quantum system is that of entanglement. When a physical system is in an entangled 
 state local measurements at one point may instantaneously affect the result of local measurements at distant points.
 A universal measure of entanglement is the Entanglement Entropy (EE). 
 It can be defined for any quantum field theory or many-body system
and unlike correlation functions is a non-local quantity. 
Suppose we are given a quantum system, e.g. a quantum field theory in $d+1$ dimensions. 
Furthermore, suppose that we split the system in two
parts, $D$ and its complementary $D^C$. Assuming that the full Hilbert space of the theory $H$ can be written as 
the direct product of the Hilbert spaces of the parts $D$ and $D^C$, namely  $H=H_D \otimes H_{D^C}$, 
one can define the reduced density matrix (RDM) of region $D$ in the following way
\begin{equation}\label{RDMgen}
\rho_D \, = \,  Tr_{D^C} \rho \, ,  
\quad \rm{where} \quad  
\rho = |\psi\rangle \langle\psi| \, , 
\end{equation}
if the system is in a pure quantum state $|\psi\rangle$.
Then the EE is defined as the von Neumann entropy of the reduced density matrix, which as can be seen from 
\eqref{RDMgen} is obtained when we trace out the degrees of freedom of the complementary region $D^C$
\begin{equation}\label{EEgen}
S_{EE}(D) \, = \,  - \, \tr{\rho_D \, \log{\rho_{D}}} \, .
\end{equation}
Physically, the EE indicates to what extend  the two subsystems, $D$ and $D^C$, are correlated. 
Equivalently, one can also interpret the EE as the entropy measured by an observer sitting in the region 
$D$ who has no access to information about the subsystem  $D^C$. 

Furthermore, since the EE is defined as the von Neumann entropy one should expect that it
is, somehow, related to the degrees of freedom of the system under consideration. 
This expectation is fully realised in the context of two dimensional conformal field theories ($2d$ CFTs) 
where the universal piece of the EE  is proportional to the central charge 
\cite{Holzhey:1994we, Calabrese:2004eu}.
Indeed, for a one-dimensional system with periodic boundary conditions at the critical point, 
the EE for an interval of length $l$ is given by  
\begin{equation}\label{Cal-Car}
S_{EE}(l) \, =  \,  \frac{c}{3} \, \log{\left(\frac{L}{\pi a} \, \sin{\frac{\pi l }{L}}\right)} 
\quad \text{with} \quad 
S_{EE}(l) \, \approx \, \frac{c}{3} \, \log{\frac{l}{a}}
\quad \rm{for} \quad  
L\rightarrow \infty
\end{equation}
where $c$ is the central charge of the corresponding CFT, $a$ is a UV cut--off and $L$ is the length of the whole system.
Based on the holographic proposal of Ryu \& Takayanagi (RT) \cite{Ryu:2006bv,Ryu:2006ef} for the EE of a higher 
dimensional quantum field theory, it seems that the proportionality between the EE and the central charge of a CFT 
in four dimensions also holds.  
The holographic proposal for calculating the EE has been proved for spherical entangling regions \cite{Casini:2011kv} and there 
are supporting arguments based on the notion of generalised entropy \cite{Lewkowycz:2013nqa}.

As opposed to the thermal entropy, the EE is non-vanishing at zero temperature. Therefore, we can employ it to probe 
the quantum properties of the ground state for a given quantum system. Additionally it can be used as an order 
parameter for the study of quantum phase transitions at zero temperature \cite{Klebanov:2007ws,Georgiou:2015pia}. 

The vast majority of the results obtained so far in the literature have been devoted to the entanglement properties 
of the vacuum state. Comparatively, very little is known about the behaviour of the EE when the system under consideration 
is in an excited state (see for example \cite{Alba:2009th})\footnote{For a review summarising the progress on the 
calculation of EE in quantum spin systems see \cite{Latorre:2009zz}.}. The aim of this work is to contribute towards 
this direction.  In particular, we will focus our attention at one of the mostly studied conformal field theories, namely the 
maximally supersymmetric field theory in four dimensions c. It is well-known that the operators of  
${\cal N}=4$ SYM can be mapped to states of an integrable spin-chain, while the dilatation operator can be mapped to a 
long-range spin-chain Hamiltonian whose eigenvalues give the spectrum of the dilatation operator \cite{Beisert:2010jr}. 
Furthermore, through the AdS/CFT correspondence \cite{Maldacena:1997re} 
(for a set of pedagogical introductions see \cite{Ramallo:2013bua, Edelstein:2009iv})
the gauge theory operators are dual to certain string states propagating on the $AdS_5 \times S^5$ 
background with the energies of the string states being equal to the dimension of the dual field theory operator.  

We should mention that the Entanglement Entropy that we are about to calculate is not directly related to the EE of 
${\cal N}=4$ SYM as a field theory. Such entropy measures the entanglement of a 3-dimensional 
subregion of the manifold on which the theory is defined to the rest of the space. What we will calculate is the EE
of a portion of the  ${\cal N}=4$ SYM spin chain, when the chain is in an excited state of either one of the closed  
rank one subsectors of  ${\cal N}=4$ SYM or in the full $PSU(2,2|4)$ algebra of  ${\cal N}=4$ SYM, in the case of the BMN limit.
The important point is that, through the AdS/CFT correspondence, the EE of the spin chain should be somehow related to 
the EE of the corresponding string state, that is to the EE of an excited state of the $1+1-$dimensional supersymmetric non-linear $\sigma-$ model 
which describes the propagation of the corresponding string in the AdS background. As it is extremely complicated to calculate this quantity 
directly from  the $\sigma-$ model considered as a field theory, 
our intension is to see if one can extract some information from the corresponding spin chain picture where the calculation is considerably 
easier\footnote{To our knowledge no precise relation between the EE,
as calculated from the $\sigma-$ model and from the corresponding spin chain, can be found in the literature.}.

The plan of the paper is as follows:
In Section \ref{EE-2magnons} we will analytically calculate the EE of excited states with two magnons in 
all closed rank one subsectors of  ${\cal N}=4$ SYM, namely $SU(2)$, $SU(1|1)$ and $SL(2)$. 
Our calculation will be performed using the formalism of Coordinate Bethe Ansatz (CBA) and will be leading in the 
coupling expansion (our states will be the eigenstates of the one-loop dilatation operators) 
but exact in the length of the spin chain and of the part of it we cut, namely $D$. 

In Section \ref{EE-BMN} we will calculate the EE of the superconformal primary operator with two excitations in the BMN limit. 
We will derive an analytic expression for the EE of which is exact in the coupling 
$\lambda'=\frac{g_{YM}^2N}{J^2}=\frac{\lambda}{J^2}$. This will allows us to analyse the effect of long-range interactions 
of the spin chain on the EE. In particular, we will see that the EE of a part of the spin chain is a monotonically increasing 
function of the coupling which saturates to a constant value as 
$\lambda' \rightarrow \infty$ when keeping the length of the chain we cut fixed. This results to a violation of a certain bound for the EE that is present at weak coupling. 
Thus, one of our main conclusions is that, as it is physically anticipated, the entanglement between parts of the 
chain becomes stronger as one increases the coupling $\lambda'$, at least for the superconformal primary operator 
with two excitations.  

In Section \ref{ABA-gen} we will employ integrability and more precisely the powerful formalism of the 
Algebraic Bethe Ansatz (ABA) in order to calculate numerically the EE of excited states with up to seven 
magnons in the $SU(2)$, $SU(1|1)$ and $SL(2)$ subsectors.
Finally, in Section \ref{Conclusions} we will present our conclusions along with directions for future research.


\section{Entanglement entropy of two magnons  in the three rank one closed subsectors of the ${\cal N}=4$ spin chain}
\label{EE-2magnons}

As discussed in the introduction, it is of the outmost importance  to calculate the 
EE for the excited states of any physical system.
This task is extremely difficult but it could provide highly non-trivial information about the physical system under consideration. 
For example, the EE can be viewed as the order parameter characterising the phase transitions which the system 
might undergo, (see e.g. \cite{Georgiou:2015pia}).

In this section, and having the AdS/CFT correspondence in mind, we will focus on the case of two 
magnons propagating in the ${\mathcal N}=4$ SYM spin chain.
In particular we will consider operators in each of the three rank one closed 
subsectors of ${\mathcal N}=4$ SYM, namely $SU(2)$, $SU(1|1)$ and $SL(2)$.
We will derive analytic expressions for any two magnon state in all the 
aforementioned sectors by employing the CBA.
 
As is well known, the problem of finding the eigenvalues and the eigenvectors of the dilatation operator of 
${\mathcal N}=4$ SYM can be solved by mapping this operator to the Hamiltonian of a certain 
integrable long-range spin chain. Then one can apply the method of Perturbative Asymptotic Bethe Ansatz (see \cite{Staudacher:2004tk} for details about this method) 
to solve for the eigenstates and the eigenvalues. If one is restricted to the one loop case, then the 
powerful technique of the ABA can be employed. 
It is this method that we will use in the following to obtain the EE for spin chains with different lengths 
and up to seven magnons.


\subsection{Entanglement Entropy of the vacuum}

We will take the operator corresponding to the vacuum state to be
\begin{equation} \label{vac}
{\cal O}_{vac} \, \sim \, \tr{Z^L} \, . 
\end{equation}
This is a BPS operator whose engineering  dimension is $L$. This dimension is not altered by quantum corrections.
The corresponding spin chain state is given by
\begin{equation}
\label{vac1}
|\downarrow\rangle_{vac}  \, = \, \, \prod_{i = 1}^L \, \otimes  \, |\downarrow\rangle_{i} \, .
\end{equation}
When the system is in the ground state \eqref{vac1} the EE of any part of the spin chain $D$ is zero, i.e. $S_{EE}(D)=0$,  
since \eqref{vac1} can be written as a direct product of states at each site.


\subsection{Entanglement Entropy of a state with one magnon}

It is straightforward to consider the case where the wavefunction of the system is that of a giant magnon with momentum $p$.
Although this is not a legitimate state since the cyclicity of the trace will necessarily set $p=0$, 
one can consider this state as a building block of states with more than one excitations.
In a spin chain language the eigenstate of a giant magnon is given by
\begin{equation} \label{magnon}
|\psi\rangle_{magnon}  \, \sim \,  \sum_{l=0}^L e^{i p l}  \, |\uparrow_{l}\rangle
\quad {\rm where} \quad 
|\uparrow_{l}\rangle \, = \, 
\prod_{i = 1}^{l-1} \, \otimes  \, 
|\downarrow\rangle_{i} \, \otimes  \, 
|\uparrow_{l}\rangle\otimes  \,\prod_{i = l+1}^{L}  \, \otimes  \, |\downarrow\rangle_{i} \, . 
\end{equation}
One can then use \eqref{magnon} to calculate the entanglement of a part of the chain with length $N$ 
to the rest of the spin chain. It is straightforward to show that the corresponding EE reads \cite{Molter:2014hna}
\begin{equation} \label{magnonEE}
S_{EE}^{1m}(N) \, = \, 
\log{\frac{L}{L \, - \, N}} \, - \, \frac{N}{L} \, \log{\frac{N}{L \, - \, N}} \, . 
\end{equation}
Notice that this expression is independent of the momentum $p$ with which the giant magnon propagates. 
In what follows we will see that the EE of any eigenstate of the one-loop Hamiltonian in the $SU(2)$, $SU(1|1)$ and 
$SL(2)$ sectors with $M$ magnons, will have as an upper bound the single magnon entropy of \eqref{magnonEE} 
multiplied by the number of magnons $M$
\begin{equation} \label{upperbound}
S_{EE}^{Mm}(N) \, \leq \, M \, S_{EE}^{1m}(N) \, .
\end{equation}


\subsection{Entanglement Entropy of a state with two magnons}

After this warm up we will now turn to the case of two magnons propagating in the  
$SU(2)$, $SU(1|1)$ and $SL(2)$ spin chains. 
The first step is to write the expression for the wavefunction in the CBA. This reads
\begin{eqnarray}\label{wave}
&& |\psi\rangle \, = \, \sum_{1 \leq x_1<x_2 \leq L}  \, \psi(x_1, x_2) \, |x_1,x_2\rangle 
\quad {\rm with}
\nonumber \\[4pt]
&& \psi(x_1, x_2) \, = \, e^{i(p_1 x_1 \, + \, p_2 x_2)} \, + \, 
S(p_2, p_1) \, e^{i(p_2 x_1 \, + \, p_1 x_2)} \, . 
\end{eqnarray}
In \eqref{wave} $L$ denotes the length of the spin chain, $x_1$ and $x_2$ the positions where 
the two magnons are sitting and $p_1$ and $p_2 \, = \, - \, p_1\, = \, - \, p$ are their momenta. 
Finally, $S(p_2, p_1)$ denotes the two-body scattering matrix in the sector under consideration. 
We will  substitute its specific value only at the end of the calculation and this will allow us to 
treat all three sectors simultaneously.

The next step consists in splitting the spin chain in two parts, one from site number 1 to site number $N$ 
which we will call part $D$ and  one from site number $N+1$ to site number $L$ which 
we will call the complementary part of $D$, namely $D^C$. 
Then one should take the trace of the complete density matrix $\rho=|\psi\rangle\langle\psi|$ 
with respect to the degrees of freedom of the complementary part
$D^C$ to obtain the reduced density matrix (RDM) corresponding to the part $D$, that is
\begin{equation}\label{red-DM-def}
\rho_D \, = \,  Tr_{{\small D^C}} \, \rho \, .
\end{equation}
In order to perform the tracing one has to distinguish three cases.


\subsection*{No magnons in the part $D$ of the spin chain}

This configuration gives the following contribution to the reduced density matrix
\begin{eqnarray}\label{red-DMi}
&& \rho_D^{(i)} \, = \,  \sum_{N < x_1<x_2 \leq L} \langle x_1,x_2|\psi \rangle \, 
\langle \psi |x_1,x_2 \rangle  \, = \, 
|\downarrow\rangle_{D\,\,\,D} \langle \downarrow| \,\,\,\,\,\, f_p(L,N) 
\quad {\rm with} 
\nonumber \\ [3pt]
&& f_p(L,N) \, = \, \sum_{N < x_1<x_2 \leq L} \, \psi(x_1, x_2) \, \psi^*(x_1, x_2)\, ,
\end{eqnarray}
where $|\downarrow\rangle_{D} $ is the vacuum for the region $D$
\begin{equation}
|\downarrow\rangle_{D}  \, = \,  \prod_{i = 1}^N \, \otimes  \, |\downarrow\rangle_{i} \, .
\end{equation}


\subsection*{Both magnons in the part $D$ of the spin chain}

The corresponding contribution to the RDM reads
\begin{equation}\label{red-DMii}
\rho_D^{(ii)} \, = \, |\psi_D\rangle \langle \psi_D |  
\qquad \rm{where} \qquad 
|\psi_D\rangle \, = \, \sum_{1 \leq x_1<x_2 \leq N}  \psi(x_1, x_2) \, |x_1,x_2\rangle  \, .
\end{equation}


\subsection*{One magnon in the part $D$ and one in the complementary $D^C$ of the spin chain}

In this case we have for the contribution to the RDM
\begin{eqnarray}\label{red-DMiii}
&& \rho_D^{(iii)} \, = \, \sum_{1 \leq x_1\leq N} \,\,\, \sum_{1 \leq x'_1\leq N} 
|x_1 \rangle \langle x'_1|\,\,\,\, g_p(x_1,x'_1) 
\quad {\rm with} 
\nonumber \\ [3pt]
&& g_p(x_1,x'_1)  \, = \, \sum_{N < x_2 \leq L}  \, \psi(x_1, x_2) \, \psi^*(x'_1, x_2)\, .
\end{eqnarray}
Combining \eqref{red-DMi}, \eqref{red-DMii} \& \eqref{red-DMiii} we finally get for the RDM
\begin{equation}\label{red-DM}
\rho_D\, = \, \kappa \, \Bigg[|\psi_D\rangle \langle \psi_D | \, + \, 
\sum_{1 \leq x_1\leq N}\,\,\, 
\sum_{1 \leq x'_1\leq N} 
|x_1 \rangle \langle x'_1|\,\,\,\, 
g_p(x_1,x'_1) \, + \, 
|\downarrow\rangle_{D\,\,\,D} \langle \downarrow|\,\,\,\, f_p\Bigg] \, , 
\end{equation}
and normalising the trace of the RDM to one we have for the constant $\kappa$
\begin{equation}\label{NORM}
Tr_D\rho_D=1 
\quad \Rightarrow \quad 
\kappa^{-1} \, = \, \langle\psi|\psi\rangle \, = \, \sum_{1 \leq x_1<x_2 \leq L} 
\psi(x_1, x_2)  \, \psi^*(x_1, x_2) \, . 
\end{equation}
It is now straightforward to write down the $\eta$-th power of the RDM
\begin{eqnarray}\label{red-DM-n}
&& \rho_D^{\eta} \, = \, \kappa^{\eta} \, \Bigg[ |\psi_D\rangle \langle \psi_D | \langle \psi_D|\psi_D \rangle^{\eta-1} \, + \, 
|\downarrow\rangle_{D\,\,\,D} \langle \downarrow| \,\, f_p^{\eta} \, + 
\nonumber \\ [3pt] 
&&  \sum_{y_1,y_2,\cdots,y_{\eta-1}\in D} |x_1 \rangle \langle x'_1|\,\,\,\, g_p(x_1,y_1)g_p(y_1,y_2)...
g_p(y_{\eta-1},x'_1) \Bigg] \, ,
\end{eqnarray}
and the only non-trivial part is in the second line of \eqref{red-DM-n}.
This can be evaluated by noticing that the structure 
$\Delta(x_1,y_1) \, = \, A_1 \, e^{i p(x_1-y_1)}\, + \, B_1^*e^{-i p(x_1+y_1)}\, + \, c.c.$, 
which is the structure of $g_p(x_1,y_1)$, maps to a similar expression
with the same spacetime structure but with different coefficients $A_1$ and $B^*_1$ 
under the following map 
\begin{equation}
R(x_1,x'_1) \, = \, \Delta(x_1,y_1) \star \Delta(y_1,x'_1) \, = \, 
\sum_{y_1\in D^C} \, \Delta(x_1,y_1) \, \Delta(y_1,x'_1) \, .
\end{equation}
Since this is the operation needed to calculate the multiple sum we obtain
\begin{equation}\label{sum}
\sum_{y_1, \ldots y_{\eta-1}\in D} |x_1 \rangle \langle x'_1|\,g_p(x_1,y_1) \ldots
g_p(y_{\eta-1},x'_1)=A_{\eta} e^{i p(x_1-x'_1)}+B_{\eta}^*e^{-i p(x_1+x'_1)}+c.c. \, ,
\end{equation}
where the coefficients $A_{\eta}$ and $B^*_{\eta}$ are given by
\begin{equation}\label{matrix}
\begin{pmatrix} 
A_{\eta}  \\
B^*_{\eta} 
\end{pmatrix}=\begin{pmatrix} 
\alpha & \beta \\
\beta^* &\alpha^* 
\end{pmatrix}^{\eta-1}
\cdot
\begin{pmatrix} 
A  \\
B^* 
\end{pmatrix}.
\end{equation}
The entries of the matrices are given by 
\begin{align}\label{ab}
A\, = \, L \, - \, N \qquad \& \qquad B \, = \, S(p_2,p_1) \, S^*_1 \, ,
\nonumber \\[4pt]
\alpha \, = \, A \, N \, + \, B \, \hat{S}_1^* \qquad \& \qquad \beta \, = \,  A \, \hat{S}_1\, + \, B N \, , 
\\[4pt]
S_1=\sum_{x \in D^C } e^{i(p_2-p_1)x} \qquad \& \qquad \hat{S}_1=\sum_{x \in D } e^{i(p_2-p_1)x} \, .
\nonumber
\end{align}
One can now diagonalise the $2 \times 2$ matrix to obtain an analytic expression for the coefficients  
$A_{\eta}$ and $B^*_{\eta}$.
Plugging this solution in the expression for the Renyi Entropy
 \begin{eqnarray}\label{Renyi-2mag}
&&S^{(\eta)}_R \, = \, \frac{1}{1\, - \, \eta} \, \log{Tr_{D} \, \rho_D^{\eta}} \qquad {\rm with}
\nonumber \\[3pt]
&&Tr_{D} \rho_D^{\eta} \, = \, \kappa^{\eta} \, 
\Bigg[ \langle \psi_D|\psi_D \rangle^{\eta} \, + \, \left(A_{\eta} \, + \, A^*_{\eta} \right) N \, + \, 
B_{\eta} \, \hat{S}_1^* \, + \, B^*_{\eta} \, \hat{S}_1 \, + \,  f_p^{\eta} \, \Bigg] \, .
\end{eqnarray}
and taking the limit $\eta \rightarrow 1$ we find for the EE of the two 
magnon excited state that
\begin{eqnarray}\label{EE-2mag}
S_{EE} & = & \lim_{\eta \rightarrow 1} S_R^{(\eta)} 
\\ [3pt]
&=&-\kappa \, \Bigg[ f_p \, \log{f_p} \, + \, \langle \psi_D|\psi_D \rangle \, \log{\langle \psi_D|\psi_D \rangle} \, + \,  
\left(\sum_{i=1}^{2} G_i \, \log{\lambda_i}  \, + \, c.c.\right) \Bigg] \, +  \, \log{\kappa} \, , 
\nonumber
\end{eqnarray}
where 
\begin{align}\label{aux-func}
& G_1 \, = \, \frac{- \, \beta^*}{2 \, \sqrt{\Delta}} \, \left(A \, - \, U_{12} \, B^*\right) 
\left(U_{11} \, N \, + \, \hat{S}_1\right) 
\quad \& \quad 
U_{11} \, = \, \frac{i \, Im \, \alpha \, - \, \sqrt{\Delta}}{\beta^*} 
\nonumber \\[4pt]
& G_2 \, = \, \frac{\beta^*}{2 \sqrt{\Delta}} \, \left(A\, - \, U_{11} \, B^*\right) 
\left(U_{12} \, N \, + \, \hat{S}_1\right)
\quad \& \quad 
U_{12} \, = \, \frac{i \, Im \, \alpha \, + \, \sqrt{\Delta}}{\beta^*}
\\[4pt]
& \lambda_1 \, = \, Re \, \alpha \, - \, \sqrt{\Delta} \, ,  \quad
\lambda_2=Re\alpha+\sqrt{\Delta}
\quad {\rm with} \quad 
\Delta \, = \, -(Im \, \alpha)^2 \, + \, |\beta|^2 \, . 
\nonumber
\end{align}
We should mention that $\lambda_1$ and $\lambda_2$ are the two eigenvalues of the $2 \times 2$ 
matrix appearing in \eqref{matrix} while $\alpha$ and $\beta$ are given in \eqref{ab}. 
A few important comments are in order. Firstly, the expression \eqref{EE-2mag} gives the EE for  all three 
closed rank one subsectors. The difference between the three 
sectors enters through the different values of the scattering matrices 
and the corresponding quantisation of the momenta. Namely, we have
\begin{align}
& S_{SU(2)}(p_1,p_2) \, = \, \frac{\cot{\frac{p_1}{2}} \, - \, \cot{\frac{p_2}{2}}\, + \, 2 i}
{\cot{\frac{p_1}{2}}\, - \, \cot{\frac{p_2}{2}}\, - \,2 i} \
\quad \Rightarrow \quad 
p_1 \, = \, - \, p_2 \, = \, \frac{2 \pi n}{L-1}
\nonumber \\[4pt]
& S_{SU(1|1)}(p_1,p_2) \, = \, - \, 1
\quad \Rightarrow \quad 
p_1 \, = \, - \, p_2 \, = \, \frac{\left(2 n \, + \, 1 \right) \, \pi}{L}
\\[4pt]
& S_{SL(2)}(p_1,p_2) \, = \, \frac{\cot{\frac{p_1}{2}} \, - \, \cot{\frac{p_2}{2}}\, - \, 2 i}
{\cot{\frac{p_1}{2}}\, - \, \cot{\frac{p_2}{2}}\, + \,2 i} \
\quad \Rightarrow \quad 
p_1 \, = \, - \, p_2 \, = \, \frac{2 \pi n}{L+1} \, . 
\nonumber
\end{align}
Notice also that the sums over $x_1$ and $x_2$ appearing in all expressions above (see equation \eqref{NORM}, for instance) should be replaced by
$\sum_{x_1\leq x_2}$ when considering the $SL(2)$ sector since in this case both derivatives may sit at the same $Z$ field.
Contrary to the common trend in the literature we define $L$ in the $SL(2)$ sector as the sum of the background fields 
plus the number of magnons.  
Secondly, we should stress that equation \eqref{EE-2mag} gives the leading contribution to the EE in the coupling expansion. 
As is well known, the ${\cal N}=4$ SYM dilatation operator and as a result its eigenvalues  receive corrections 
order by order in perturbation theory. These corrections will also affect the value of the EE. Our calculation, in this 
section, gives the leading term in the weak coupling expansion of the EE. 
However, it is exact  as a function of the spin chain length. 

\begin{figure}[h] 
   \centering
   \includegraphics[width=7.5cm]{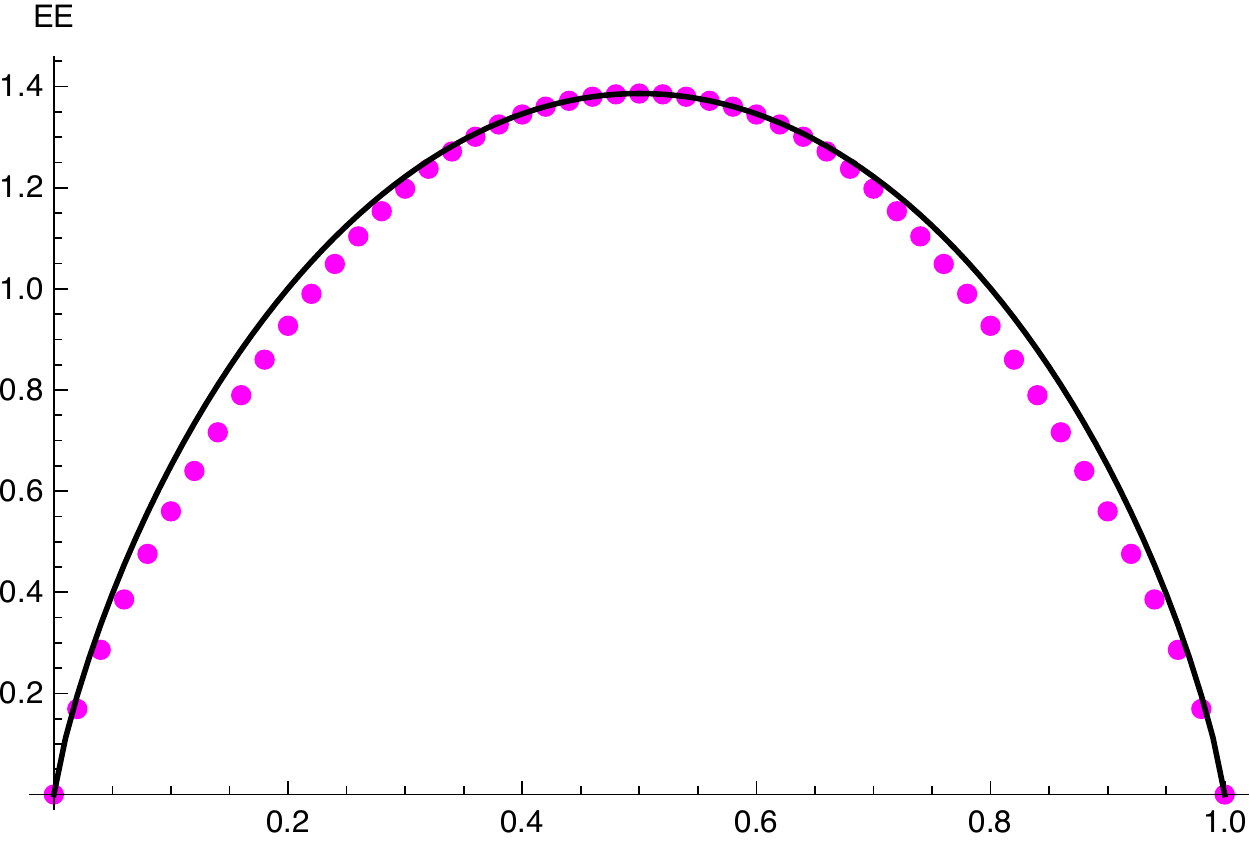}
    \includegraphics[width=7.5cm]{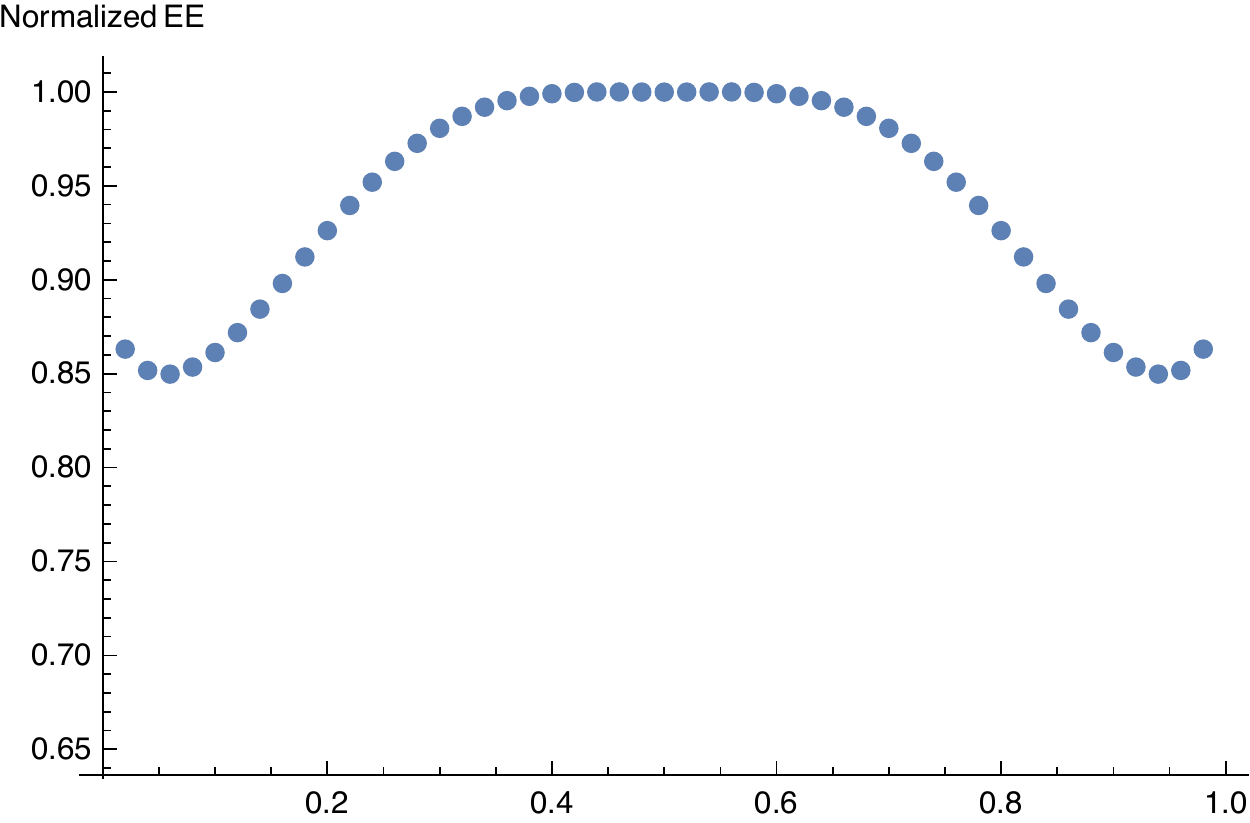}
       \caption{In this figure we present the EE for an excited state with two magnons in the $SU(2)$ sector. 
       On the left part of the figure it is the plot of the EE as a function of the position that we split the spin chain 
       in two parts, $D$ and its complement $D^C$. In order to simplify notation and to be able to compare spin chains 
       with different number of sites we normalize the horizontal axis and plot with respect to the ratio of the 
       splitting point divided by the length of the spin chain. 
       In all the subsequent plots of the EE the horizontal axis will be in units of this ``normalized splitting" ($N/L$). 
       The black curve is twice the EE of a single magnon while the magenta dots represent the actual 
       computation of the EE, using \eqref{EE-2mag}, when the scattering of the two magnons is taken into account.
       On the right part of the figure, in order to illustrate the saturation points, we present the normalized EE 
       (i.e. dividing \eqref{EE-2mag} by twice the EE of a single magnon). The calculations are for mode number $n=1$ 
       \& the length of the spin chain is set to $L=100$.}
   \label{fig:1.1}
\end{figure}

Given the general expressions for the Renyi Entropy  \eqref{Renyi-2mag} and the EE \eqref{EE-2mag}, 
we are now in position to plot the EE as a function of the size of the part of the spin chain we cut. In figure \ref{fig:1.1} 
we present the EE for an excited state with two magnons in the $SU(2)$ sector. 
The quantum number specifying the momenta of the magnons is set to $n=1$.
On the left part of the figure one can see the plot of the EE as a function of the position that we split the spin chain 
in two parts, $D$ and its complement $D^C$. In the same plot we have also drawn twice the 
EE of a single magnon, which is the upper bound for the EE of any state involving two magnons. 
One can see that the bound is almost saturated at two symmetric points, one on the left and one on the right 
of the middle of the chain (see also the right part of Figure \ref{fig:3.1})\footnote{This is not the case for the $SL(2)$ 
subsector where the bound (for $n=1$) is almost saturated when we cut the spin chain in two equal parts, see figure 
\ref{fig:3.3}.}. To illustrate this point we present the normalised EE,
that is the ratio of \eqref{EE-2mag} over twice the EE of the single magnon, on the right part of the figure \ref{fig:1.1}.
\begin{figure}[h] 
   \centering
   \includegraphics[width=7.5cm]{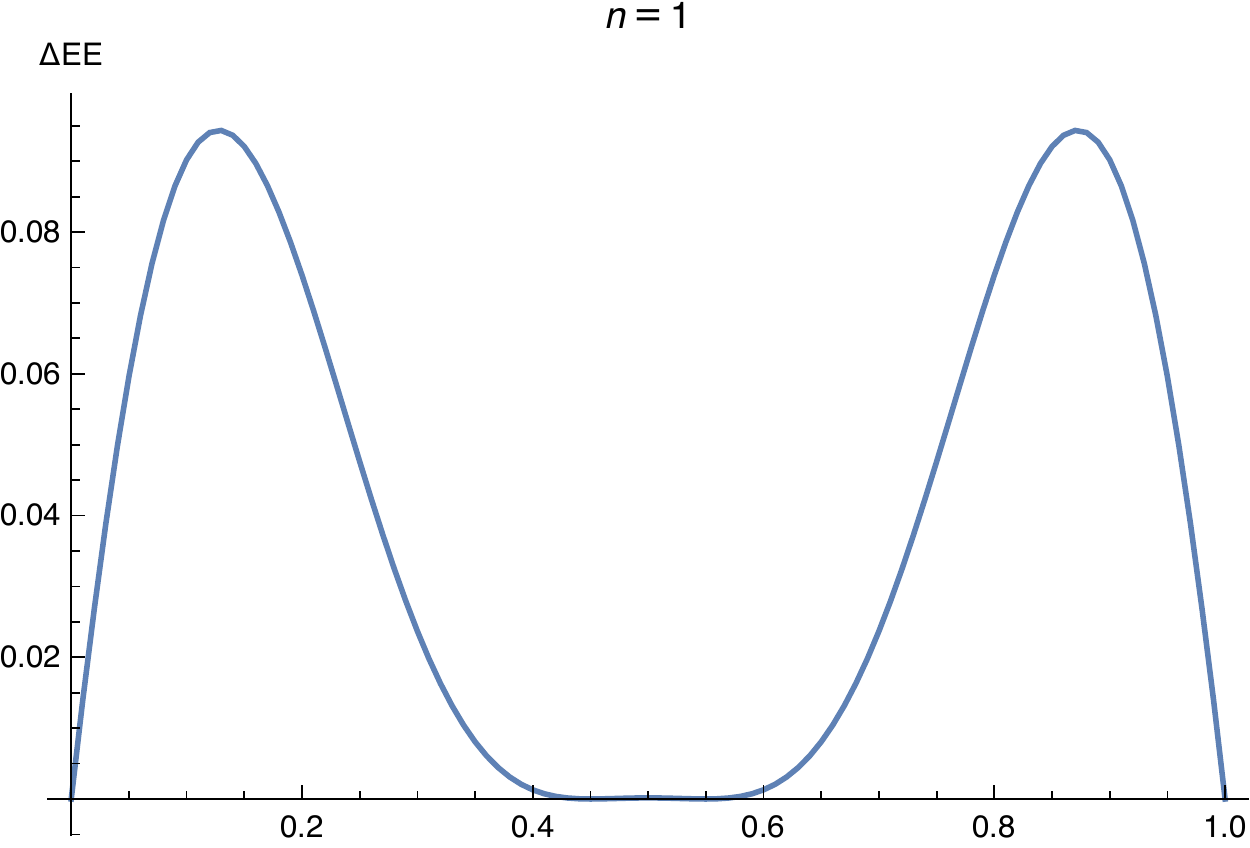}
    \includegraphics[width=7.5cm]{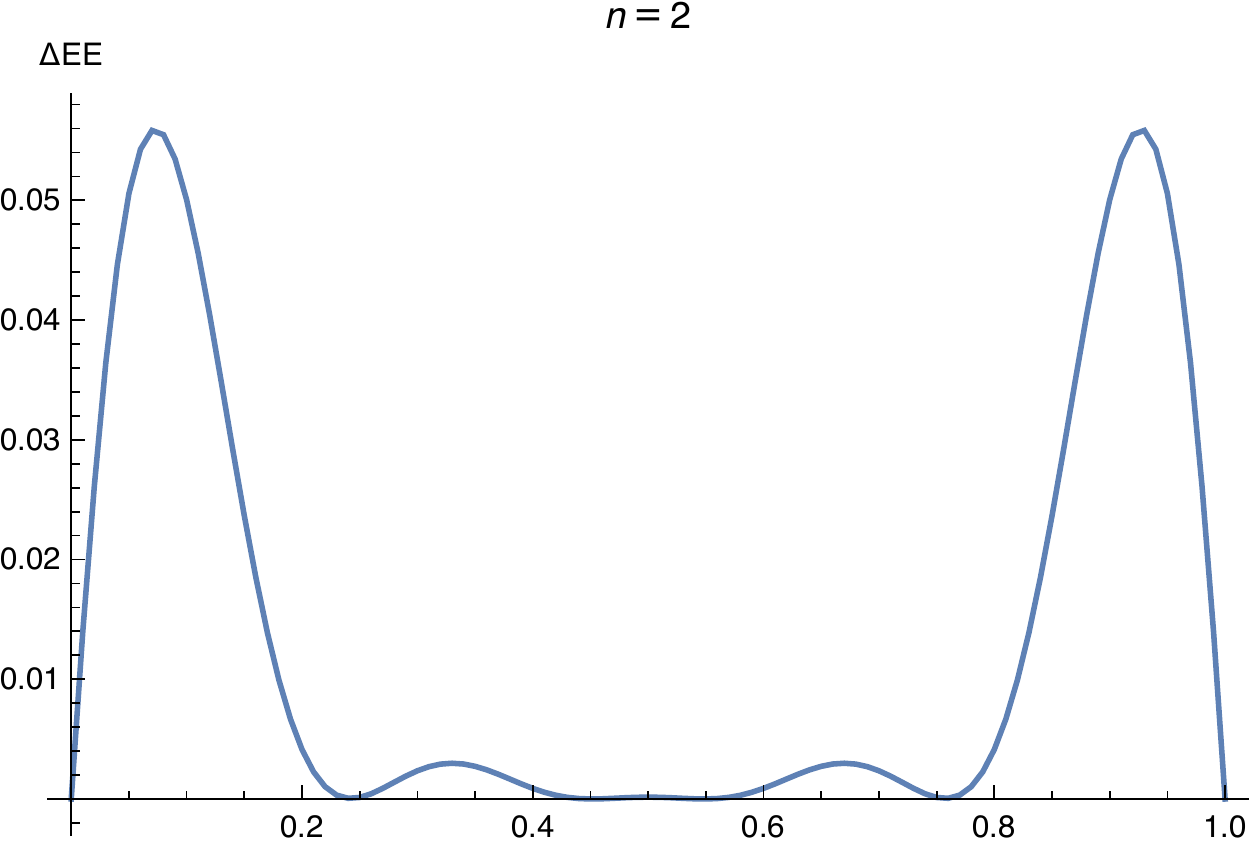}
    \includegraphics[width=7.5cm]{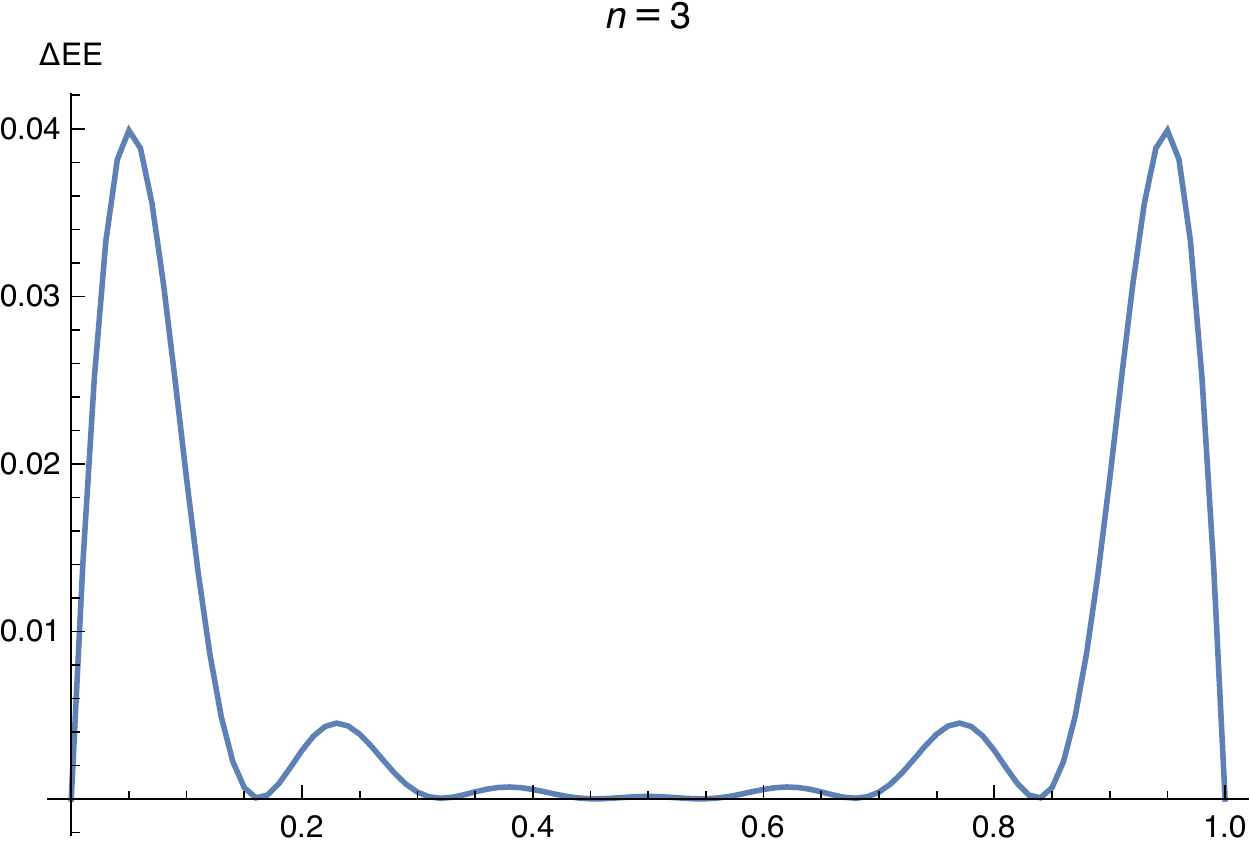}
     \includegraphics[width=7.5cm]{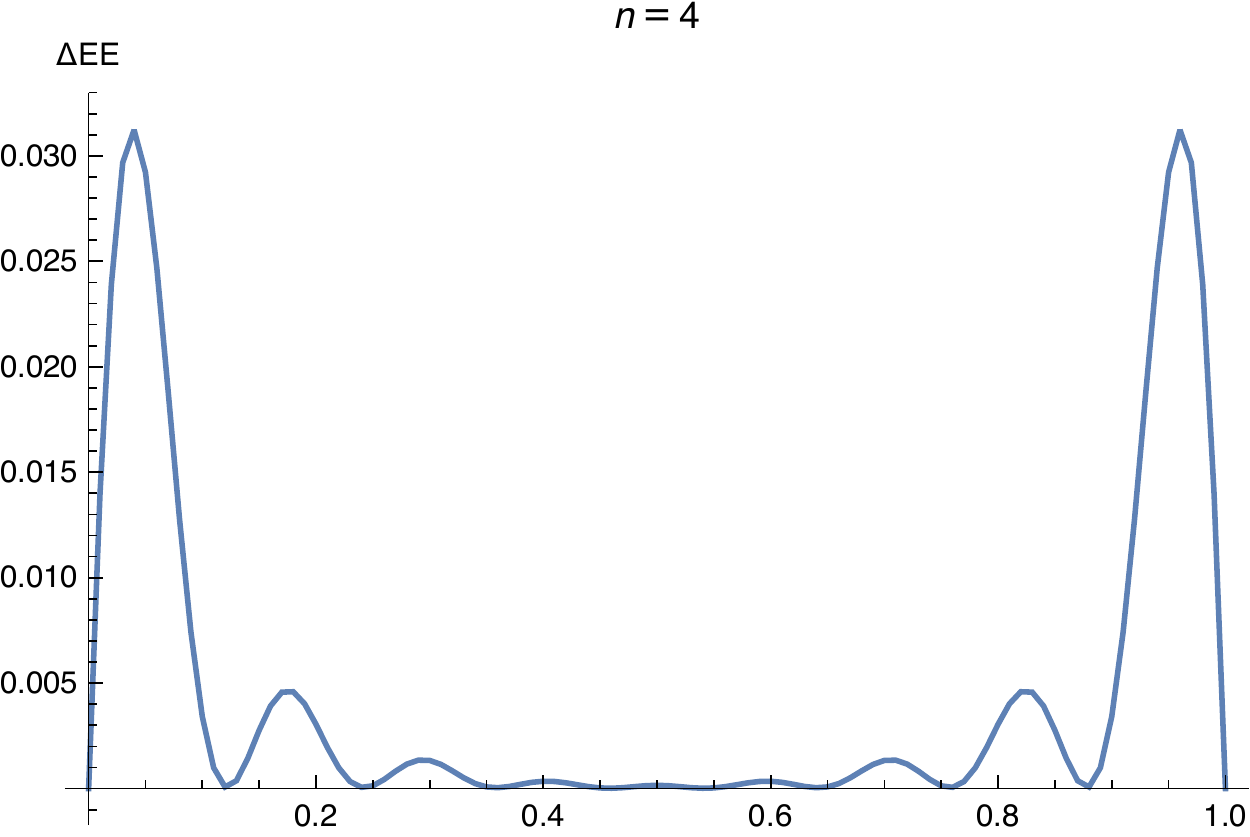}
       \caption{In this figure we present the difference between the EE of two magnons and twice the EE of one magnon 
       when we change/increase the mode number (from mode number one to four), in the $SU(2)$ sector.}
   \label{fig:1.2}
\end{figure}
In figure \ref{fig:1.2} we present the dependence of the difference between the aforementioned bound and the EE of two 
magnon excited state on the mode number $n$, which characterises the excited state. Generically, as one moves 
towards the centre of the chain the difference oscillates with an amplitude that decreases. 
Furthermore, as one increases the mode number from one to four the bound is almost saturated for specific values 
of the length of the domain $D$ for which the EE is calculated. If we exclude the trivial cases where $D$ 
is either the empty set or when $D$ is the whole chain the number of points that the bound is almost saturated is $2n$
(see also the end of the next paragraph), which is twice the excitation number. 
We should note that part of the results of the current section have some overlap with the analysis in \cite{Molter:2014hna}.

\begin{figure}[h] 
   \centering
   \includegraphics[width=7.5cm]{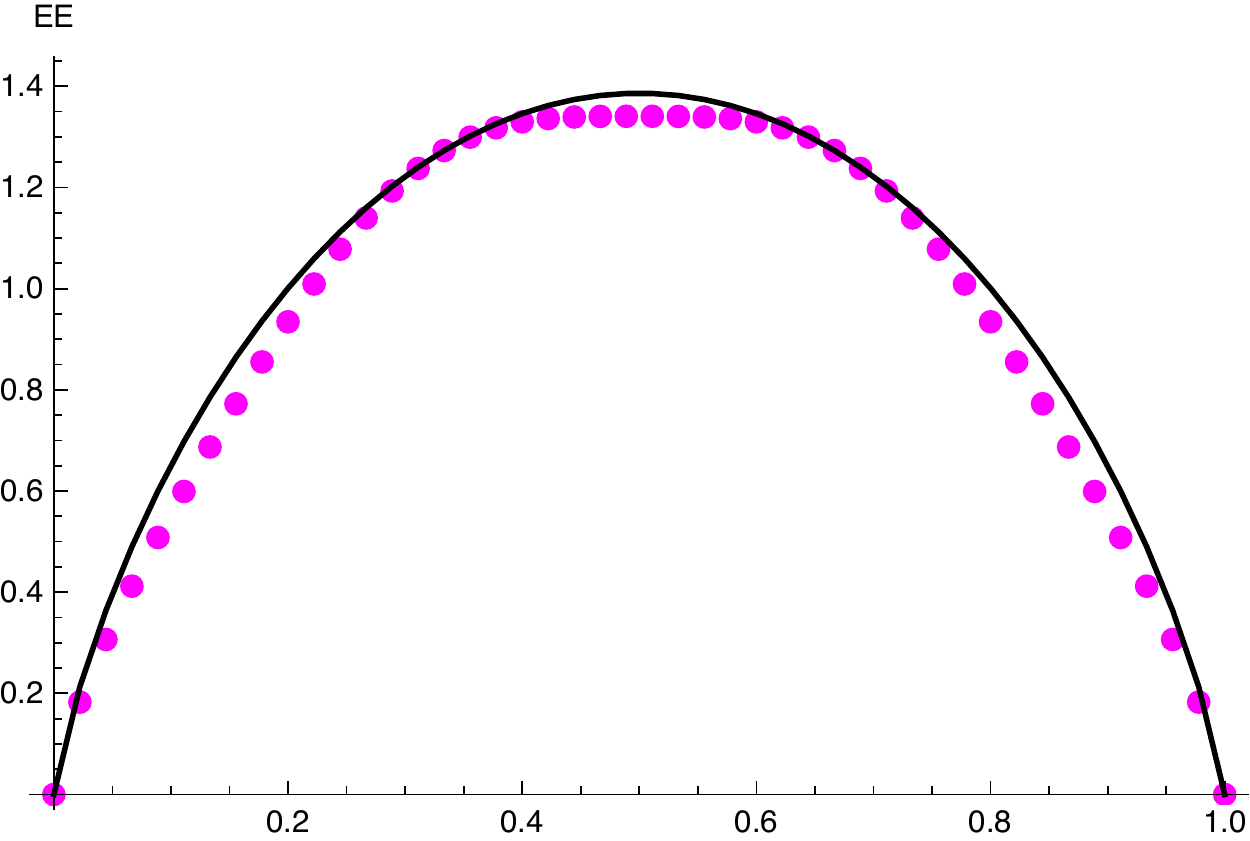}
    \includegraphics[width=7.5cm]{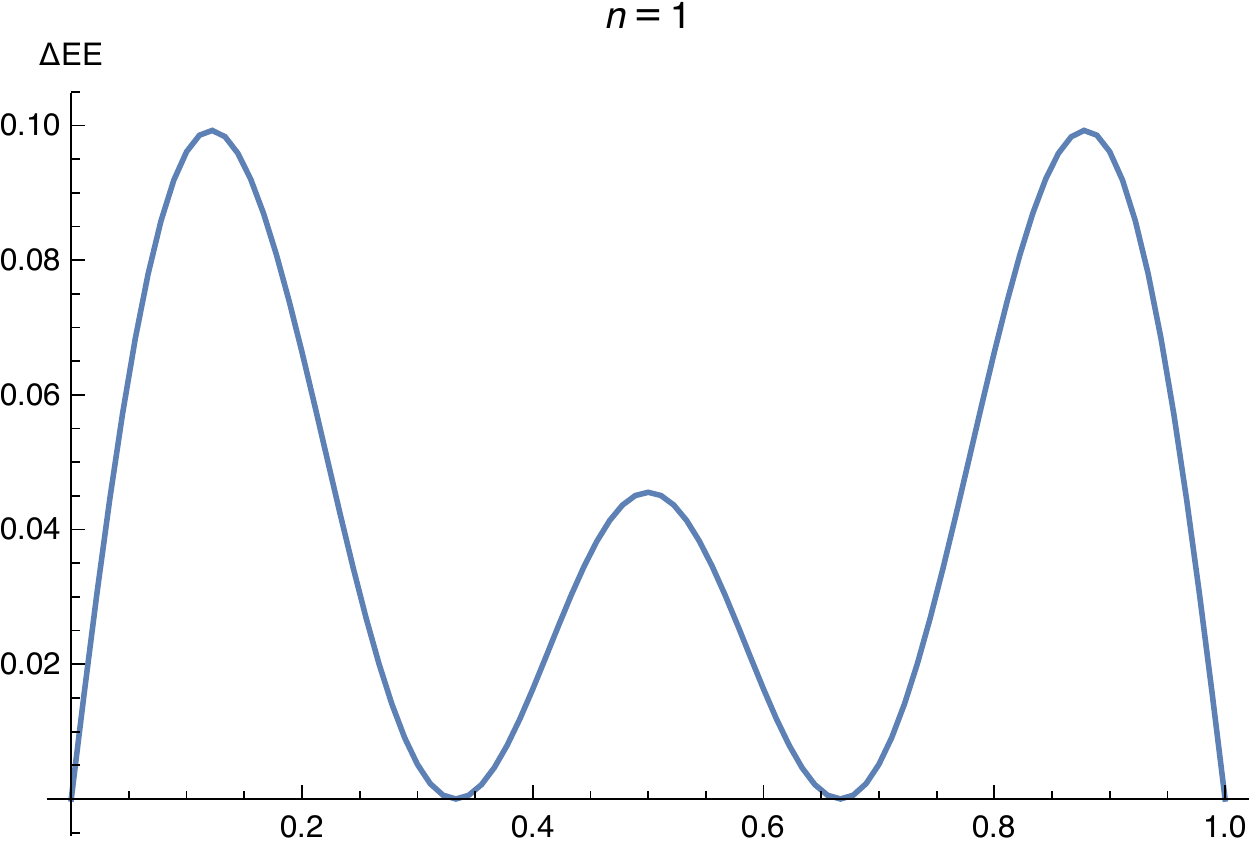}
    \includegraphics[width=7.5cm]{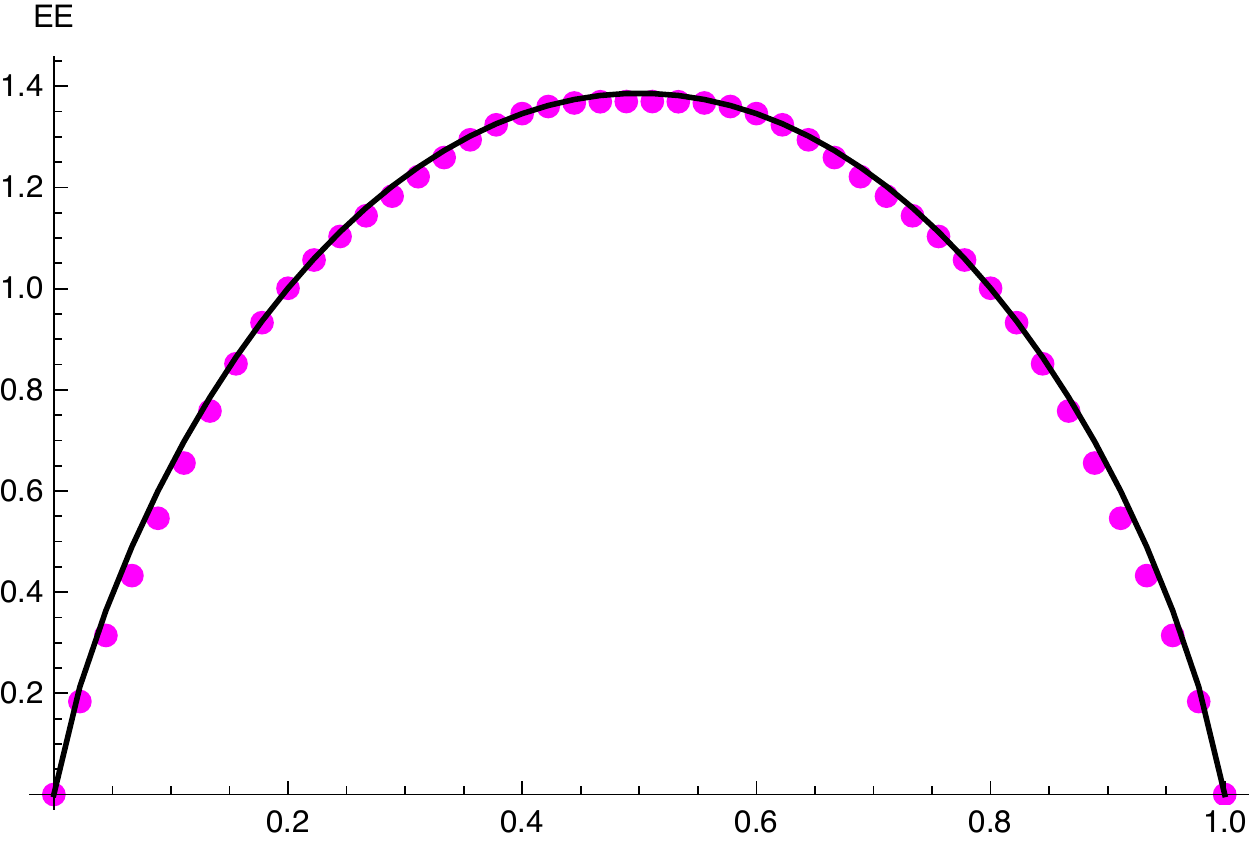}
     \includegraphics[width=7.5cm]{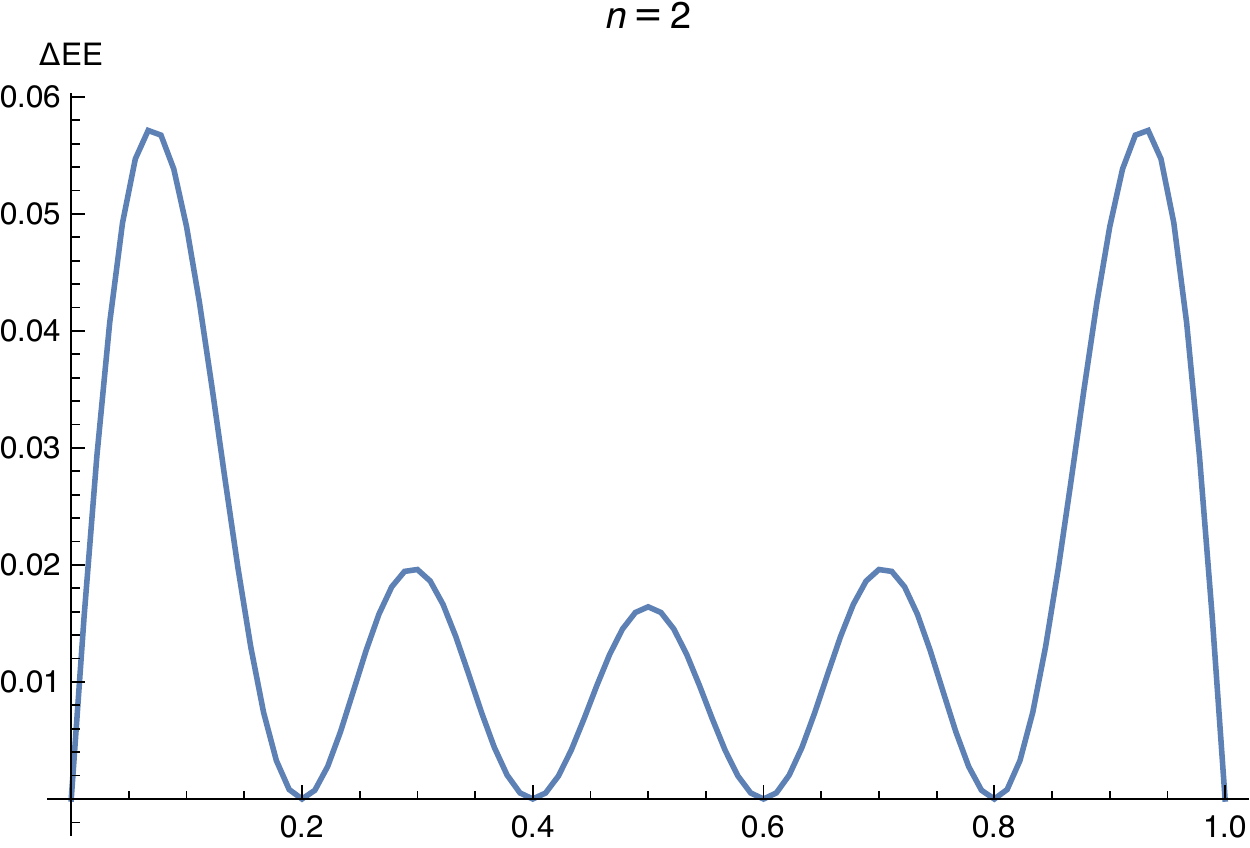}
       \caption{In this figure we present the EE for an excited state with two magnons in the $SU(1|1)$ sector. 
       On the upper left part of the figure it is the plot of the EE as a function of the ``normalised splitting" ($N/L$) 
       and on the upper right part the difference between the EE of two magnons and twice the EE of the single magnon again
       as a function of the ``normalised splitting", when the mode number is $n=1$. In the lower two plots of the figure we 
       present the results for mode number $n=2$, to illustrate the increase of the number of the explicit saturation 
       points when the mode number increases.   
       The black curve is twice the EE of a single magnon while the magenta dots represent the
       computation of the EE using \eqref{EE-2mag}. The length of the spin chain is set to $L=90$.}
   \label{fig:1.3}
\end{figure}

Looking at the expression of the EE bound (see \eqref{magnonEE} multiplied by two) it is clear that if there points 
where this bound is explicitly saturated, then the analytic expression of the EE \eqref{EE-2mag}  at those points
should be independent of the value of the momentum of the excitation. To detect those points, one should go to
\eqref{ab} where all the ingredients are defined and look for a systematic way to eliminate the presence of the 
momentum. Setting $S_1$ and $\hat{S}_1$ to zero satisfies the above requirement and furthermore 
eliminates the presence of the momentum in the expression of $f_p$ and $\psi_D$. In that way the EE 
at those points, which simultaneously set to zero $S_1$ and $\hat{S}_1$, explicitly saturate the bound \eqref{upperbound}.
Combining the expressions for $S_1$ and $\hat{S}_1$ together with the value of the momentum for 
each one of the three sectors, it is easy to conclude that the bound is explicitly saturated only in the $SU(1|1)$
sector and in the following points 
\begin{equation}\label{saturation-points}
\frac{N}{L} \, = \,   \frac{\kappa}{2 n +1} \quad  {\rm with} \quad 
n \ge 0 \quad \& \quad 0 \le \kappa \le 2n+1.
\end{equation}
To illustrate the above claim we have plotted in figure \ref{fig:1.3} the EE in the $SU(1|1)$ sector for two 
different quantum numbers, namely $n=1$ and $n=2$, in a spin chain with 90 sites. 
According to \eqref{saturation-points} we expect to 
have four and six saturation points, located at the positions $(0,1/3,2/3,1)$ and $(0,1/5,2/5, 3/5,4/5, 1)$ respectively, 
and this is exactly what we observe in figure \ref{fig:1.3}. 
In the other two sectors the quantities $S_1$ and $\hat{S}_1$ can never be simultaneously zero and for that reason 
the entropy comes very close to the bound but without saturating it. 
In cases with more than two magnons the existence
of saturation points seems difficult to occur, since more than two 
constraints have to be satisfied simultaneously, but needs to be checked with an explicit calculation. 

 A final comment concerns the expected fact that the plots for the EE are symmetric 
with respect to the centre of the spin chain. This is a consequence of the well-known fact that the EE's of $D$ 
and of its complementary $D^C$ are equal $S_{EE}(D)=S_{EE}(D^C)$, when the
state which describes the system as a whole $D \cup D^C$ is a pure state.

The aim of the next section will be to analyse the effect of interactions to the EE. To this end we will focus on 
the BMN limit of $AdS_5 \times S^5$ and find the exact, in the coupling, expression for the EE. 


\section{All-loop Entanglement and Renyi Entropies of the superconformal primary operator  with two excitations in the BMN limit}
\label{EE-BMN}

One of the most interesting limits of the AdS/CFT correspondence is the so-called BMN limit \cite{Berenstein:2002jq}. 
The reason is that in this limit, known as the Penrose or pp-wave limit on the gravity side, 
the Green-Schwarz superstring action for type IIB strings becomes 
quadratic in the light-cone variables and as a result one can solve for the superstring spectrum exactly.
This result provides an all-orders prediction for the anomalous dimensions of certain operators with large 
R-charge \cite{Beisert:2005tm} which are dual to the string states propagating in the pp-wave background.
Subsequently, one can study the dynamics of the theory by employing string field theory 
to construct the three-string vertex (see \cite{Lee:2004cq,Dobashi:2004nm,Chu:2002pd} and references therein) 
and compare the so-obtained string amplitudes to the corresponding three-point correlators 
\cite{Georgiou:2003aa,Georgiou:2004ty,Georgiou:2009tp}\footnote{For weak/strong coupling comparisons  of three-point correlators see also \cite{Costa:2010rz,Georgiou:2010an,Georgiou:2011qk}.}.

 
\subsection{The superconformal primary operator with two impurities in the BMN limit}
\label{BMN}

In this subsection, we will briefly review the  construction of the superconformal primary state involving 
two excitations (impurities) 
on  both the string and gauge theory side. The full supermultiplet based on this primary state was constructed in
\cite{Beisert:2002tn} at leading order in the coupling expansion.
In what follows, we will closely follow \cite{Georgiou:2008vk,Georgiou:2009tp}.
The main idea is to use the action of the superalgebra on the states of the theory in order to resolve the operator mixing 
appearing in the wavefunction  of the primary operator. More precisely, consider the non-BPS highest weight state (HWS) 
with two impurities which we will denote by ${\cal O}_n$.
By definition this state should be annihilated by the sixteen superconformal generators preserved by the pp-wave background. 
Schematically one has
\begin{equation}\label{si}
[S,{\cal O}_n(x=0)]=0~~~~~\mbox{and}~~~~~ [Q,{\cal O}_n(x=0)]\not=0~. 
\end{equation}
The action of the remaining sixteen supersymmetry generators, collectively denoted here by $Q$, 
on the HWS generates the whole supermultiplet with two impurities. 
Equation \eqref{si} should be implemented order by order in perturbation theory since the 
superconformal charges receive quantum corrections \cite{Georgiou:2008vk,Georgiou:2009tp}.
However in the pp-wave limit the 32 supercharges can be straightforwardly constructed order by order in the string coupling.
What is important for us is that their leading in $g_s$ expressions are known 
to all-orders in the effective Yang-Mills coupling $\lambda'=\frac{g_{YM}^2 N}{J^2}$.
Here $g_{YM}^2 N$ is the 't Hooft coupling while $J$ is the large R-charge of the operator which correspond to the 
angular momentum of the point-like string orbiting around one of the equators of the five-sphere $S^5$ 
of the parent $AdS_5 \times S^5$ background.
Demanding that the full set of the 16 superconformal charges annihilates the HWS one can determine the form of the latter to all orders in $\lambda'$. The details of this construction can be found in  \cite{Georgiou:2008vk}. The result for the two-impurity HWS reads
\begin{equation}
\label{shws}
|n\rangle = \frac{1}{4(1+U_n^2)}
\Bigg[
    {a^\dagger}_n^{i'}
    {a^\dagger}_n^{i'}
    \,+\,
    {a^\dagger}_{-n}^{i'}
    {a^\dagger}_{-n}^{i'}
  + 2 U_n 
  b_{-n}^\dagger\, \Pi\; b_n^\dagger
  - U_n^2 \left(
    {a^\dagger}_n^{i}
    {a^\dagger}_n^{i}
    \,+\,
    {a^\dagger}_{-n}^{i}
    {a^\dagger}_{-n}^{i}
  \right)\Bigg]
  |\alpha\rangle \, .  
\end{equation}
In \eqref{shws} ${a^\dagger}_{\pm n}^{i'}$, $ b_{\pm n}^\dagger$ and ${a^\dagger}_{\pm n}^{i}$ 
denote the creation operators for the four scalar, eight fermionic and four vector excitations
while $ |\alpha\rangle$ denotes the string vacuum of fixed light-cone
momentum $p^+$. 
Furthermore, $n$ is the mode number characterising the excited state while 
the function $U_n$ is given by 
\begin{equation}
U_n \equiv \frac{1-\rho_{n}}{1+\rho_{n}}
\quad {\rm with} \quad 
\rho_{n} \, = \, \frac{\omega_n \, - \, n}{\mu \, \alpha}
 \quad \& \quad 
\omega_{n} \, = \, \sqrt{n^2 \, + \, \left(\mu \, \alpha' \, p^+\right)^2} \, , 
\label{cppm}
\end{equation}
where  $p^+$ is the light cone momentum of the state and $\mu$ is the parameter setting
the scale of the curvature of the PP-wave background 
(as usual, $\alpha \equiv \alpha' p^+$ and $\lambda'=1/(\mu \alpha)^2$).
Finally, $\Pi$  is the appropriate $16\times 16$ block of the matrix $\Pi=\prod_{i'=1}^4\Gamma^{i'}$.
The index $i'$  takes values in the flavour $SO(4) \subset SO(6)$ and the $\Gamma$  
indicate the $SO(1,9)$ gamma matrices.

One important comment is in order. Notice that the HWS string state, as well as the corresponding 
field theory operator ${\cal O}_n$, exhibits 
the important feature of mixing between different kinds of excitations, namely bosonic and fermionic states (operators) 
mix among each other as long as the mixing states have the same quantum numbers.

Needless to say that this construction can be generalised to HWS with more than two impurities.


\subsection{Exact in $\lambda'$ Entanglement Entropy}
\label{exact-BMN}

Our aim in this section is to derive, based on \eqref{shws}, an analytic expression for the EE of the two impurity primary 
operator which is exact in the BMN coupling $\lambda'$. This expression will be an interpolating function 
from the weak coupling regime $\lambda'\rightarrow 0$ to the strong coupling 
regime $\lambda'\rightarrow\infty$. We should stress that our result is exact in the strict BMN limit and generically 
will receive $1/J$ corrections. It would be interesting to calculate these corrections by going to the near-BMN limit.

The first step towards this end is to rewrite \eqref{shws} as an operator of ${\cal N}=4$ SYM. 
An important observation is that due to mixing of  different kinds of impurities this operator can not be restricted in one of the 
closed subsectors of ${\cal N}=4$ SYM but "lives" in the full $PSU(2,2|4)$ superalgebra.
The field theory operator which is dual to the string state  \eqref{shws} can be written as follows
\begin{align} \label{primary}
(1+U_n^2){(\mathcal{O}_{st})}^J_n 
& =
\sqrt{\frac{N_0^{-J-2}}{J+3}} \, \sum_{i=1}^2 \sum_{p=0}^J \, \cos{\frac{\pi n(2p+3)}{J+3}} \, 
\tr{Z_i \, Z^p \, \bar{Z}_i \, Z^{J-p}}
\nonumber \\
& - 
2\, \sqrt{\frac{N_0^{-J-2}}{J+3}} \cos{ \frac{\pi n}{J+3}} \, \tr{\bar{Z} \, Z^{J+1}}
\\[4pt]
\qquad \qquad \qquad \qquad \qquad& +
\frac{U_n}{2} \, \sqrt{\frac{N_0^{-J-1}}{J+1}} \,  \sum_{p=0}^{J-1} \sin{\frac{\pi n(2p+2)}{J+1}} \, 
 \tr{\psi^{1 \alpha} \,Z^p  \, \psi^2_{\alpha} \, Z^{J-p-1}}
\nonumber \\[4pt]
&  -
\frac{U_n}{2} \, \sqrt{\frac{N_0^{-J-1}}{J+1}} \, \sum_{p=0}^{J-1} \sin{\frac{\pi n(2p+2)}{J+1}} \,   
\tr{\bar\psi_{3 \dot\alpha} \, Z^p \, \bar\psi_4^{\dot\alpha} \, Z^{J-p-1}} 
\nonumber \\[4pt]
& \,-\, 
\frac{U_n^2}{4} \, \sqrt{\frac{N_0^{-J}}{J-1}} \sum_{p=0}^{J-2} \cos{\frac{\pi n(2p+1)}{J-1}} \, 
\tr{D_{\mu}Z \, Z^p \, D^{\mu} \, Z \, Z^{J-p-2}} \, .
\nonumber
\end{align}
In \eqref{primary} $\psi^A$ (with $A=1,2,3,4$) denote the four fermions of ${\cal N}=4$ SYM while $\alpha=1,2$ and $\dot \alpha=1,2$ are spinor indices over which we sum. Furthermore, $N_0=\frac{N}{8 \pi^2}$ where $N$ is the number of colours.
We should mention that in order to translate the string state \eqref{shws} to the field theory operator 
\eqref{primary} we have used the prescription of \cite{Georgiou:2008vk}. 
Namely, we have used the following dictionary
\begin{equation} \label{eq:a1}
\frac{1}{4}\left[ {a^\dagger}_n^{i'} \, {a^\dagger}_n^{i'} \,+\, 
{a^\dagger}_{-n}^{i'} \, {a^\dagger}_{-n}^{i'} \right]| \, \alpha\rangle
 \,\leftrightarrow\, 
 \mathcal{O}^{(0) J}_n \, , 
\end{equation}
where $\mathcal{O}^{(0) J}_n$ is the sum of the first two terms on the r.h.s. of \eqref{primary}. 
The string state on the l.h.s. of \eqref{eq:a1} is normalised to one and the same is true for  the tree-level
2-point function of the corresponding gauge theory operator. 
In a similar fashion the term with the fermionic oscillators in \eqref{shws} corresponds to
a field theory operator with four fermions
\begin{equation} \label{eq:a2}
\frac{1}{2} \left[ {b^\dagger}_{-n} \, \Pi \,{b^\dagger}_n \right]| \, \alpha\rangle
\,\leftrightarrow\,
\mathcal{O}^{(1) J}_n \, , 
\end{equation}
where $\mathcal{O}^{(1) J}_n$ is the sum of the third and fourth terms on the r.h.s. of \eqref{primary}
multiplied by $U_n$. Finally, for the term involving the vector impurities we have
\begin{equation} \label{eq:a3}
\frac{1}{4}\left[ {a^\dagger}_n^{i} \, {a^\dagger}_n^{i} \,+\, 
 {a^\dagger}_{-n}^{i} \, {a^\dagger}_{-n}^{i} \right]| \, \alpha\rangle
 \,\leftrightarrow\,
 \mathcal{O}^{(2) J}_n \, , 
\end{equation}
where $ \mathcal{O}^{(2) J}_n $ is the last term on the r.h.s. of \eqref{primary} multiplied by $U_n^2$.
As in the purely scalar operator, \eqref{eq:a2} and \eqref{eq:a3} are derived so that both the string state 
and the gauge theory operator are normalised to one. Since the two-point functions of operators involving 
fermions or/and vector impurities have non-trivial space-time structure
we have used the prescription of \cite{Georgiou:2004ty,Georgiou:2003kt}. 
Once is given an operator it is possible to define the barred one, which is the conjugate of the initial 
operator followed by an inversion. It is then this operator which is used in the calculation of the two-point function.
This prescription is motivated by the radial quantisation in two-dimensional CFT's and 
results to two-point functions which can be easily normalised to one. 

One can now use the following relation
\begin{eqnarray}\label{ident}
&& \sum_{p=0}^J\cos{\frac{\pi n(2p+3)}{J+3}} \, 
\tr{X \, Z^p \, Y \, Z^{J-p}} \, \approx \, \frac{1}{2 \, J}   \times 
\\ [4pt]
&& \sum_{1 \leq x_1<x_2 \leq J} 
\left(e^{i (p_1 x_2 \, + \, p_2 x_1)} \, + \,  e^{i (p_1 x_1 \, + p_2 x_2)}\right) \, 
\tr{Z^{x_1-1} \, X \, Z^{x_2-x_1-1} \, Y \, Z^{J-x_2}} \, ,
\nonumber
\end{eqnarray}
as well as the analogous equation for the fermionic term to rewrite \eqref{primary} as a spin chain wavefunction.
To achieve this one should take into account the fact that $p_1=p=-p_2=2 \pi n/J<<1$ and as a result to leading order in the 
large $J$ expansion $e^{\pm i p} \approx 1$ and $\cos\big({ \pi n/(J+3)}\big)\approx 1$. 
Furthermore, we will make use of the following 
correspondence between the Yang-Mills and spin chain excitations \cite{Georgiou:2012zj}
\begin{eqnarray}\label{corresp}
\frac{1}{\sqrt{N_0}} \,  Z_{YM} \leftrightarrow Z_{sp} 
\qquad 
\frac{1}{\sqrt{2 \,N_0}} \, \psi^A_{YM} \leftrightarrow \psi^A_{sp} 
\qquad 
\frac{1}{\sqrt{2 \, N_0}} \, \left(D_{\mu}Z\right)_{YM} \leftrightarrow \left(D_{\mu}Z\right)_{sp} \, .
\end{eqnarray}
In conclusion the wavefunction of the non-BPS primary operator with two excitations can be written in the spin chain language as (we suppress the index "sp" since from now on we will be using only the spin chain wavefunctions) follows
\begin{eqnarray}\label{spinstate}
|\psi\rangle & =& 
\frac{\mathcal N}{2 J}\, \Bigg[ \sum_{i=1}^2 \sum_{x_1<x_2}
\left(e^{i p_1 x_2+i p_2 x_1}+ e^{i p_1 x_1+i p_2 x_2}\right) 
|(Z_i)_{x_1}({\bar Z}_i)_{x_2}\rangle \, - \, 
4\sum_{x}|{\bar Z}_{x}\rangle 
\\ [4pt]
&-& i \, U_n \sum_{x_1<x_2}\left(e^{i p_1 x_2+i p_2 x_1}- e^{i p_1 x_1+i p_2 x_2}\right)
\left(|(\psi^{1\alpha})_{x_1}(\psi^{2}_{\alpha})_{x_2}\rangle - 
|({\bar \psi}_{3 \dot \alpha})_{x_1}({\bar \psi}_{4}^{\dot \alpha})_{x_2}\rangle \right)
\nonumber \\[4pt]&-& \frac{U_n^2}{2} \sum_{x_1<x_2}
\left(e^{i p_1 x_2+i p_2 x_1}+ e^{i p_1 x_1+i p_2 x_2}\right)
|(D_{\mu}Z)_{x_1}(D^{\mu}Z)_{x_2}\rangle \Bigg] \, . 
\nonumber 
\end{eqnarray}
A couple of important comments are in order. As is stressed in \cite{Beisert:2005tm} the ${\cal N}=4$ SYM spin 
chain is dynamic
in the sense that the length of the chain in not fixed since different impurities have different scaling dimensions. 
Indeed as can be seen from \eqref{primary} the number of $Z$ fields is not the same in all terms. 
We should mention that each of the kets written in \eqref{spinstate} describe two excitations in the appropriate number of 
$Z$ fields dictated by \eqref{primary}. The dynamic nature of the spin chain may result to a situation
where both vector impurities lie in the domain $D$ but the scalar impurities lie on in $D$ and one in its complementary. 
This can happen when both excitations are close to the boundary of  $D$. 
In what follows we will ignore such circumstances since their contribution will be $1/J$ suppressed in the BMN limit.
As second related comment concerns the exact form of \eqref{primary}. As is well-known the exact expression for the 
eigenstate of the dilatation operator will have non-asymptotic terms where the two impurities will be close to each other. As it 
happens with the two-point functions one can show that these terms give a contribution which is also $1/J$ suppressed with  
respect to the contributions coming from the asymptotic terms and as such can be ignored in the strict BMN limit where 
$J\rightarrow \infty$. Finally, let us notice that the state \eqref{primary} may, in principle, be taken from considering the 
scattering of two scalar impurities with a double copy of the $SU(2|2)$ scattering matrix of \cite{Beisert:2005tm}.

We are now in position to write down the RDM originating from the wavefunction above. 
As usual we will  be cutting the spin chain 
into two parts. One part is from site 1 to site $N$, which is the domain $D$ of which the EE we intend to calculate, 
while the remaining part is the complementary $D^C$ whose degrees of freedom we have to trace out in order 
to obtain the RDM. Furthermore, as in the case of two magnons at weak coupling (see section \ref{EE-2magnons}) 
by $|\psi_D\rangle$ we denote the wavefunction of the part $D$ when both magnons
sit in the region $D$, while by $|\psi_D^C\rangle$ we denote the wavefunction of the complementary region 
$D^C$ when both magnons sit in the region $D^c$. After these explanations the RDM can be written as
\begin{eqnarray} \label{red-DM-exact}
\rho_D & =&  \kappa \, \Bigg[|\psi_D\rangle \langle \psi_D | \, + \, 
|\downarrow\rangle_{D\,\,\,D} \langle \downarrow|\,\,\,\, f_p \, 
+ 4^2 \left(\sum_{x} |(\bar Z)_{x} \rangle  \langle (\bar Z)_{x}| \, + \, 
|\downarrow\rangle_{D\,\,\,D} \langle \downarrow| (J-N) \right)
\nonumber \\ [5pt]
&+& \sum_{x_1 < x'_1}\, 
\Bigg(
\sum_{i=1}^{2}  |(Z_i)_{x_1} \rangle \langle (Z_i)_{x'_1}|\,\,\,\,  g_p^{(1)}(x_1,x'_1) +
|(\psi^{1\alpha})_{x_1} \rangle \langle (\psi^{1\alpha})_{x'_1}|\,\,\,\, 
g_p^{(2)}(x_1,x'_1) 
\nonumber \\ [5pt]
&+&
|(\psi_{3\dot \alpha})_{x_1} \rangle \langle (\psi_{3\dot \alpha})_{x'_1}|\,\,\,\, 
g_p^{(2)}(x_1,x'_1) +|(D_{\mu}Z)_{x_1}\rangle  \langle D_{\mu}Z)_{x'_1}|\,g_p^{(3)}(x_1,x'_1) 
\Bigg) \Bigg] \, , 
\end{eqnarray}
where 
\begin{align}
& g_p^{(1)} \, = \, g_p^{SU(2)}
\quad {\rm with} \quad 
S_{SU(2)}(p_2,p_1) \, = \, 1 \, ,
\nonumber \\[4pt]
& g_p^{(2)}=U_n^2 g_p^{SU(1|1)}
\quad {\rm with} \quad 
S_{SU(1|1)}(p_2,p_1) \, = \, -\, 1 \, , 
\\[4pt]
& g_p^{(3)}=\frac{U_n^4}{4}g_p^{SU(2)}
\quad {\rm with} \quad 
S_{SL(2)}(p_2,p_1) \, = \, 1 \,  ,
\nonumber
\end{align}
that is $g_p^{(1)}$ is the same function appearing in the weak coupling calculation of 
section \ref{EE-2magnons} but with the scattering matrix set to one and so on.
We should mention that the fact that the scattering matrices should be set to $\pm 1$ 
is in accordance with the standard lore that in the  BMN limit the impurities do not scatter 
at all since they are most of the time very far from each other.  

It is now straightforward to write down the n-th power of the RDM
\begin{align}\label{red-DM-nexact}
\rho_D^{\eta} & =  \, \kappa^{\eta} \, 
\Bigg[ |\psi_D\rangle \langle \psi_D | \langle \psi_D|\psi_D \rangle^{\eta-1} \, + \, 
|\downarrow\rangle_{D\,\,\,D} \langle \downarrow| \,\, f_p^{\eta} 
\nonumber \\[4pt]
&+ \, 
4^{2\eta} \left(N^{\eta-1}\sum_{x}|(\bar Z)_{x}\rangle  \langle (\bar Z)_{x}| \, + \, 
\left(J \, - \, N\right)^{\eta} \, |\downarrow\rangle_{D\,\,\,D} \langle \downarrow| \right)
\\[4pt] 
&+ \, \sum_{x_1 < x'_1}\,
\Bigg(\sum_{i=1}^{2}
|(Z_i)_{x_1} \rangle \langle (Z_i)_{x'_1}|\,\,
\sum_{y_1,...,y_{\eta-1}\in D}g_p^{(1)}(x_1,y_1)g_p^{(1)}(y_1,y_2)...g_p^{(1)}(y_{\eta-1},x'_1)
\nonumber \\[4pt]
&+ \, |(\psi^{1\alpha})_{x_1} \rangle \langle (\psi^{1\alpha})_{x'_1}|\,\,\,\, 
\sum_{y_1,...,y_{\eta-1}\in D}g_p^{(2)}(x_1,y_1)g_p^{(2)}(y_1,y_2)...
g_p^{(2)}(y_{\eta-1},x'_1) 
\nonumber \\[4pt]
&+ \, |(\psi_{3\dot \alpha})_{x_1} \rangle \langle (\psi_{3\dot \alpha})_{x'_1}|\,\,\,\, 
\sum_{y_1,...,y_{\eta-1}\in D}g_p^{(2)}(x_1,y_1)g_p^{(2)}(y_1,y_2)...
g_p^{(2)}(y_{\eta-1},x'_1)
\nonumber \\[4pt]
&+ \, |(D_{\mu}Z)_{x_1}\rangle  \langle D_{\mu}Z)_{x'_1}|\,
\sum_{y_1,...,y_{\eta-1}\in D}g_p^{(3)}(x_1,y_1)g_p^{(3)}(y_1,y_2)...
g_p^{(3)}(y_{\eta-1},x'_1) \Bigg) \Bigg] \, .
\nonumber
\end{align}
A couple of comments are in order. The first term in the first line of \eqref{red-DM-nexact} 
originates from the partition where both magnons 
are in the region $D$, while the second term from the partition where both magnons are in the complementary region $D^C$.
Furthermore, the second line of  \eqref{red-DM-nexact} comes from part of the wavefunction 
which has a single impurity $\bar Z$.
Finally, the rest of the expression originates from the partition where 
one of the magnons is in the region $D$ and one in the complementary $D^C$.

The next step is to find the exact in $\lambda'$ Renyi Entropy
\begin{eqnarray}\label{Renyi-2mag-exact}
S^{(\eta)}_R &= & \frac{1}{1\, - \, \eta} \, \log{Tr_{D} \, \rho_D^{\eta}} \qquad {\rm with}
\\[4pt]
Tr_{D} \rho_D^{\eta} \, &=& \, \kappa^{\eta} \, 
\Bigg[ \langle \psi_D|\psi_D \rangle^{\eta} +  f_p^{\eta} +
\left(2^{\eta} +\left(U_n^{4}\right)^{\eta}\right) \, 
\left[\left(A^{(1)} + A^{(1)*} \right) N + B^{(1)} \, \hat{S}_1^*+ B^{(1)*} \hat{S}_1 \right]
\nonumber \\ [4pt]
 &+&\left(4U_n^2\right)^{\eta} \, 
 \left[\left(A^{(2)} +  A^{(2)*} \right) N + B^{(2)} \hat{S}_1^* + B^{(2)*} \hat{S}_1\right] + 
 4^{2\eta}\left( N^{\eta} + \left(J - N \right)^{\eta} \right) \Bigg] \, ,
 \nonumber
\end{eqnarray}
where all quantities are defined in section \ref{EE-2magnons}.
We should only add that $A^{(1)}=A^{SU(2)}$ and $B^{(1)}=B^{SU(2)}$ are the $SU(2)$ 
weak coupling expressions for $A$ and $B$ 
defined in section \ref{EE-2magnons} but with $SU(2)$ scattering matrix set to one, i.e. $S_{SU(2)}(p_2,p_1)=1$.
Similarly, $A^{(2)}=A^{SU(1|1)}$ and $B^{(2)}=B^{SU(1|1)}$ are the $SU(1|1)$ weak coupling expressions for $A$ and $B$ 
defined in section \ref{EE-2magnons} but with $SU(1|1)$ scattering matrix set to minus one, i.e. $S_{SU(1|1)}(p_2,p_1)=-1$.
Before we continue with the $\eta \rightarrow 1$ limit and the calculation of the EE, we would like to 
comment on the coefficients of the different contributions in \eqref{Renyi-2mag-exact} (namely scalar, fermion and vector) 
and their $\eta$ dependence.  

The second line of \eqref{red-DM-nexact} gives the last term in \eqref{Renyi-2mag-exact}.
The third line of \eqref{red-DM-nexact} gives the part of the third term in \eqref{Renyi-2mag-exact} which is proportional to $2^{\eta}$. This happens since $i=1,2$
so we have to raise two to the $\eta$-th power, i.e $2^{\eta}$.
The fourth and fifth lines of \eqref{red-DM-nexact} give the penultimate term in \eqref{Renyi-2mag-exact}. 
Since $\alpha=1,2$ and ${\dot \alpha}=1,2$ we have to multiply $U_n$ by four and then raise 
to the $\eta$-th power, i.e $\left(4\times U_n^2\right)^{\eta}$. In the same fashion, since $\mu=1,2,3,4$, 
the ultimate term in \eqref{red-DM-nexact} has to be multiplied by four and then raised 
to the $\eta$-th power, i.e. $\left(4\times \frac{U_n^4}{4} \right)^{\eta}$ to give the part of the third term of 
\eqref{Renyi-2mag-exact} that is proportional to $U_n^{4\eta}$. 

Taking the limit $\eta \rightarrow 1$ we find for the EE of the two  magnon excited state
\begin{align}\label{EE-2mag-exact}
S_{EE}  &= \, 
- \, \kappa \, \Bigg[ 16 \Big[N \log{(16 N)} \, + \,  \left(J-N\right) \, \log{\big( 16\left(J-N\right)\big)}\Big] \,+ \, 
f_p \, \log{f_p}  
\nonumber \\ 
 &+ \, 
 \langle \psi_D|\psi_D \rangle \, \log{\langle \psi_D|\psi_D \rangle} \, + \, \Bigg(
 4 U_n^2 \sum_{i=1}^{2}  G_i^{(2)} \, \log{(4 \, U_n^2 \, \lambda_i^{(2)})} 
\\[4pt]
 &+ \, 
2 \sum_{i=1}^{2} G_i^{(1)} \, \log{(2 \, \lambda_i^{(1)})} \, + \, 
U_n^4 \sum_{i=1}^{2} G_i^{(1)} \, \log{(U_n^4 \, \lambda_i^{(1)})} \, + \, c.c. \Bigg) \Bigg] \, + \,  \log{\kappa} \, , 
\nonumber
\end{align}
where in full analogy with \eqref{aux-func}
\begin{eqnarray}\label{aux-func-exact}
&& G_1^{(i)} \, = \, \frac{- \, \beta^{(i)*}}{2 \, \sqrt{\Delta^{(i)}}} \, \left(A^{(i)} \, - \, U_{12}^{(i)} \, B^{(i)*}\right) 
\left(U_{11} ^{(i)}\, N \, + \, \hat{S}_1\right) 
\,\, \& \quad 
U_{11}^{(i)} \, = \, \frac{i \, Im \, \alpha^{(i)} \, - \, \sqrt{\Delta^{(i)}}}{\beta^{(i)*}} 
\nonumber \\ [4pt]
&& G_2^{(i)} \, = \, \frac{\beta^{(i)*}}{2 \sqrt{\Delta^{(i)}}} \, \left(A^{(i)}\, - \, U_{11}^{(i)} \, B^{(i)*}\right) 
\left(U_{12}^{(i)} \, N \, + \, \hat{S}_1\right)
\,\, \& \quad 
U_{12}^{(i)} \, = \, \frac{i \, Im \, \alpha^{(i)} \, + \, \sqrt{\Delta^{(i)}}}{\beta^{(i)*}}
\\[4pt]
&& \lambda_1^{(i)} \, = \, Re \, \alpha^{(i)} \, - \, \sqrt{\Delta^{(i)}} \, ,  \quad
\lambda_2^{(i)}=Re\alpha^{(i)}+\sqrt{\Delta^{(i)}}
\quad {\rm with} \quad 
\Delta^{(i)} \, = \, -(Im \, \alpha^{(i)})^2 \, + \, |\beta^{(i)}|^2 \, .
\nonumber
\end{eqnarray}
The upper index $i=1$ denotes that the quantities that carry it are defined in the $SU(2)$ sector 
with the scattering matrix set to one.
Similarly,  when the upper index $i$ is set to two, $i=2$,  denotes quantities as defined in the 
$SU(1|1)$ sector with the scattering matrix set to minus one. 
Finally, the expressions for the inner products in \eqref{Renyi-2mag-exact} \& \eqref{EE-2mag-exact} are
\begin{eqnarray}\label{inner-prod}
&& \langle \psi|\psi \rangle \, = \,  2 \langle \psi|\psi \rangle^{SU(2)} \, + \, 16 J \, + \, 
4 U_n^2 \langle \psi|\psi \rangle^{SU(1|1)} \, + \, U_n^4\langle \psi|\psi \rangle^{SU(2)} 
\nonumber \\ [7pt]
 && \langle \psi_D|\psi_D \rangle \, = \, 2  \langle \psi_D|\psi_D \rangle^{SU(2)} \, + \, 
 4 U_n^2 \langle \psi_D|\psi_D \rangle^{SU(1|1)} \, + \, U_n^4 \langle \psi_D|\psi_D \rangle^{SU(2)}
 \\ [7pt]
 && f_p \, = \, 2  \langle \psi_{D^C}|\psi_{D^C} \rangle^{SU(2)} \, + \, 
 4 U_n^2 \langle \psi_{D^C}|\psi_{D^C} \rangle^{SU(1|1)} \, + \, U_n^4 \langle \psi_{D^C}|\psi_{D^C} \rangle^{SU(2)} \, , 
\nonumber
\end{eqnarray}
where $\langle \psi|\psi \rangle^{SU(2)}$ and $\langle \psi|\psi \rangle^{SU(1|1)}$ denote the inner product for the eigenstate of the one-loop dilatation operator in the $SU(2)$ and $SU(1|1)$ sectors, respectively. These are defined in section \ref{EE-2magnons}.

\begin{figure}[h] 
   \centering
   \includegraphics[width=7.5cm]{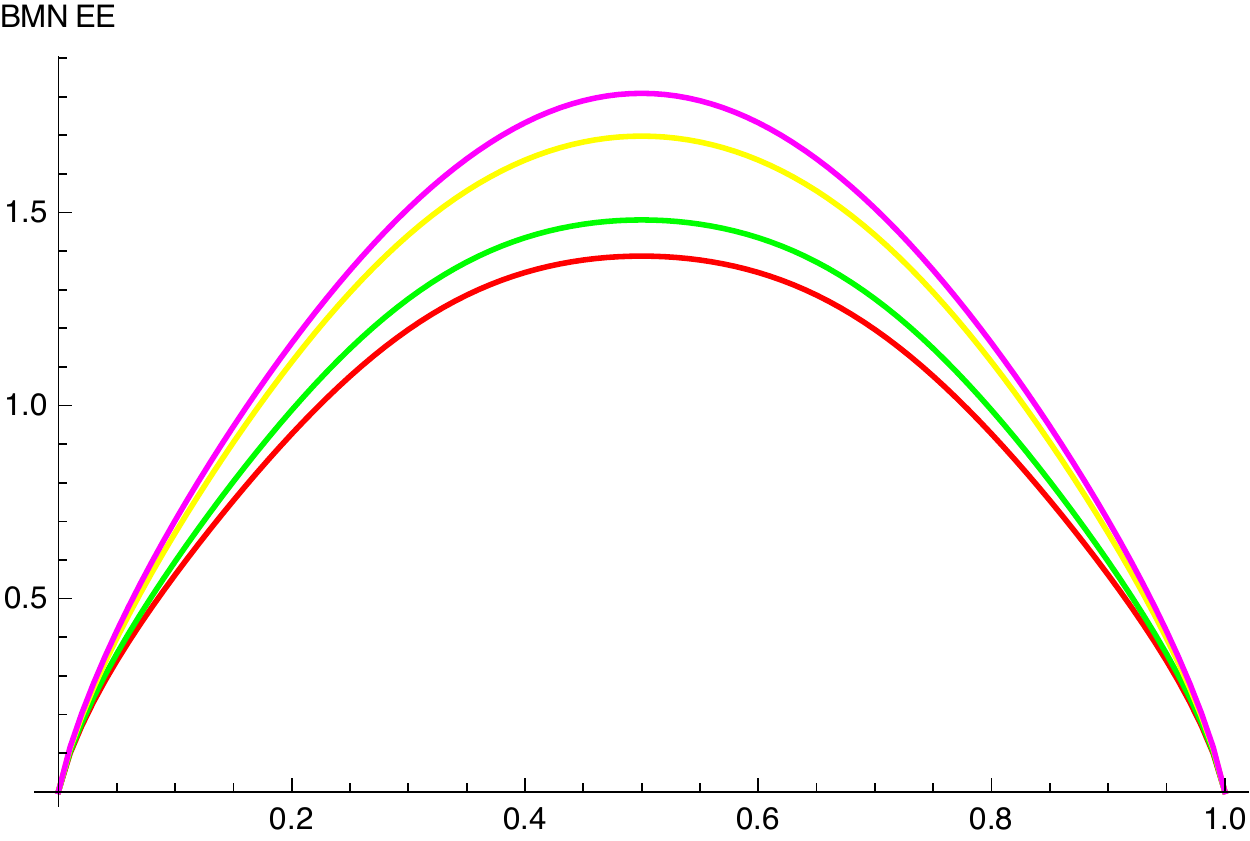}
    \includegraphics[width=7.5cm]{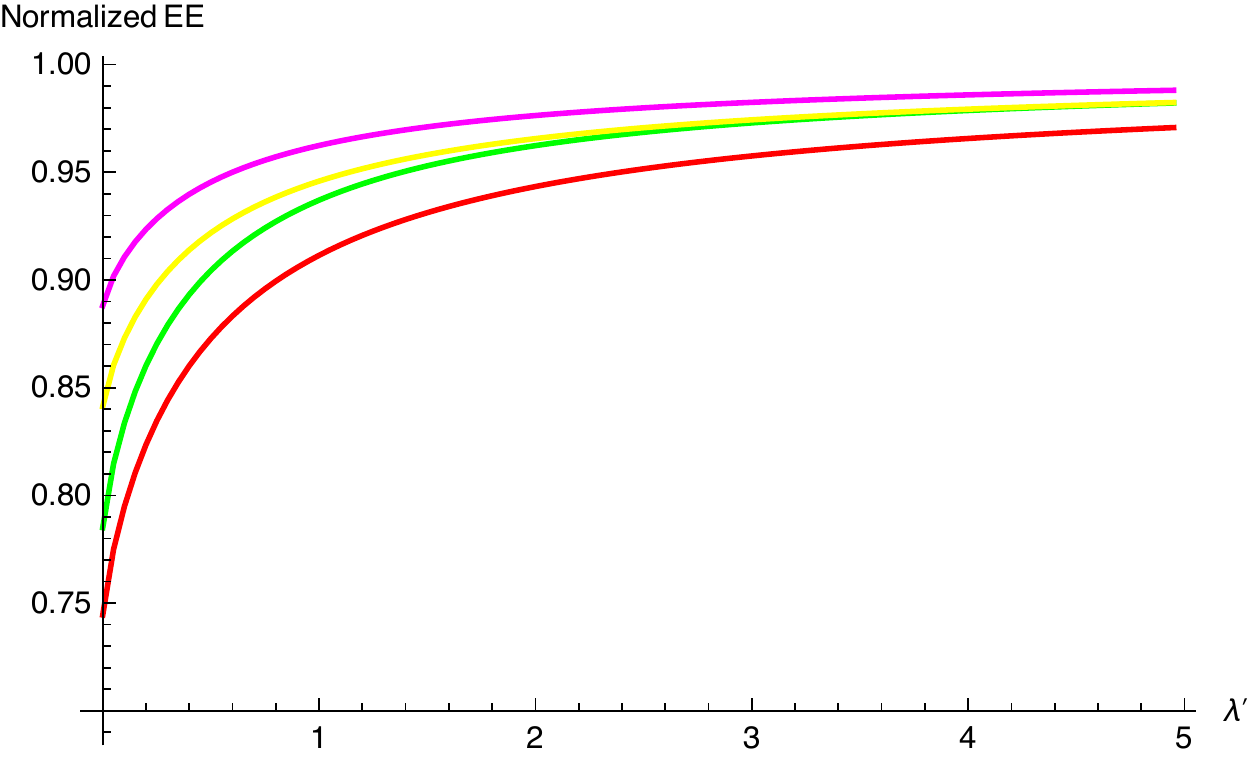}
       \caption{In this figure we present the EE for an excited state with two magnons in the BMN limit. The mode number of the state is taken to be
       $n=1$. On the left part of the 
       figure it is the plot of the EE as a function of the normalised splitting of the spin chain, for different values of $\lambda'$.
       The correspondence between colour and $\lambda'$ is the following: Red $\Rightarrow$ $\lambda'=0$, 
       Green $\Rightarrow$ $\lambda'=0.1$,  Yellow $\Rightarrow$ $\lambda'=1$ \& Magenta $\Rightarrow$ $\lambda'=5$.
       On the right part of the figure we present the EE flow from the IR (lower values of $\lambda'$) to the UV (higher
       values of $\lambda'$) as we change the splitting point of the spin chain. In order to compare the different curves we 
       have normalised each one by dividing with the EE for $\lambda'\rightarrow \infty$. In this way all the curves 
       approximate to one. The correspondence between colours and normalised splitting (e.g. $N/L$) 
       of the spin chain is the following:  Red $\Rightarrow$ $0.5$, Green $\Rightarrow$ $0.1$,  
       Yellow $\Rightarrow$ $0.01$ \& Magenta $\Rightarrow$ $0.001$.}
   \label{fig:2.1}
\end{figure}

Now having at hand the analytic expressions for both the Renyi entropy \eqref{Renyi-2mag-exact} and the EE 
\eqref{EE-2mag-exact}, we will probe the parametric space and extract interesting qualitative behaviours. 
As can be seen from both \eqref{Renyi-2mag-exact} \& \eqref{EE-2mag-exact} both quantities depend on the position 
we split the spin chain in two parts ($D$ and its complement  $D^C$), the coupling constant $\lambda'$ and the mode number
$n$ characterising the excited state\footnote{Renyi entropy depends also on the order $\eta$.}. 
In figure \ref{fig:2.1} we present the EE for an excited state with $n=1$  in the BMN limit.
On the left part of the figure it is the plot of the EE as a function of the normalised splitting of the spin chain, 
for different values of $\lambda'$. As can be seen from the plot (the correspondence colour/$\lambda'$ is explained in the 
caption of the figure) the EE increases as we increase the value of the coupling.  
To fully realise/visualise  this EE flow from the UV to IR, on the right part of figure \ref{fig:2.1} 
we present the EE as a function of the coupling $\lambda'$, for different values of the splitting point of the spin chain. 
In order to compare the different curves in a unified manner we have normalised each one by dividing with the 
EE for $\lambda'\rightarrow \infty$, i.e. 
\begin{equation} \label{normal-def}
{\rm Normalised} \, S_{EE}(N,\lambda',n) \, = \, \frac{S_{EE}(N,\lambda',n)}{S_{EE}(N,\infty,n)} \,.
\end{equation} 
Since the EE is related to the central charge of the underlying CFT, this monotonically decreasing behaviour of the 
EE along the RG flow from $\lambda'\rightarrow\infty$ to $\lambda'\rightarrow 0$  
could be related to the existence of a $c$-theorem \cite{Myers:2010xs, Myers:2010tj}, which connects through an 
RG flow a fixed point in the UV with another fixed point in the IR. 

\begin{figure}[h] 
   \centering
   \includegraphics[width=7.5cm]{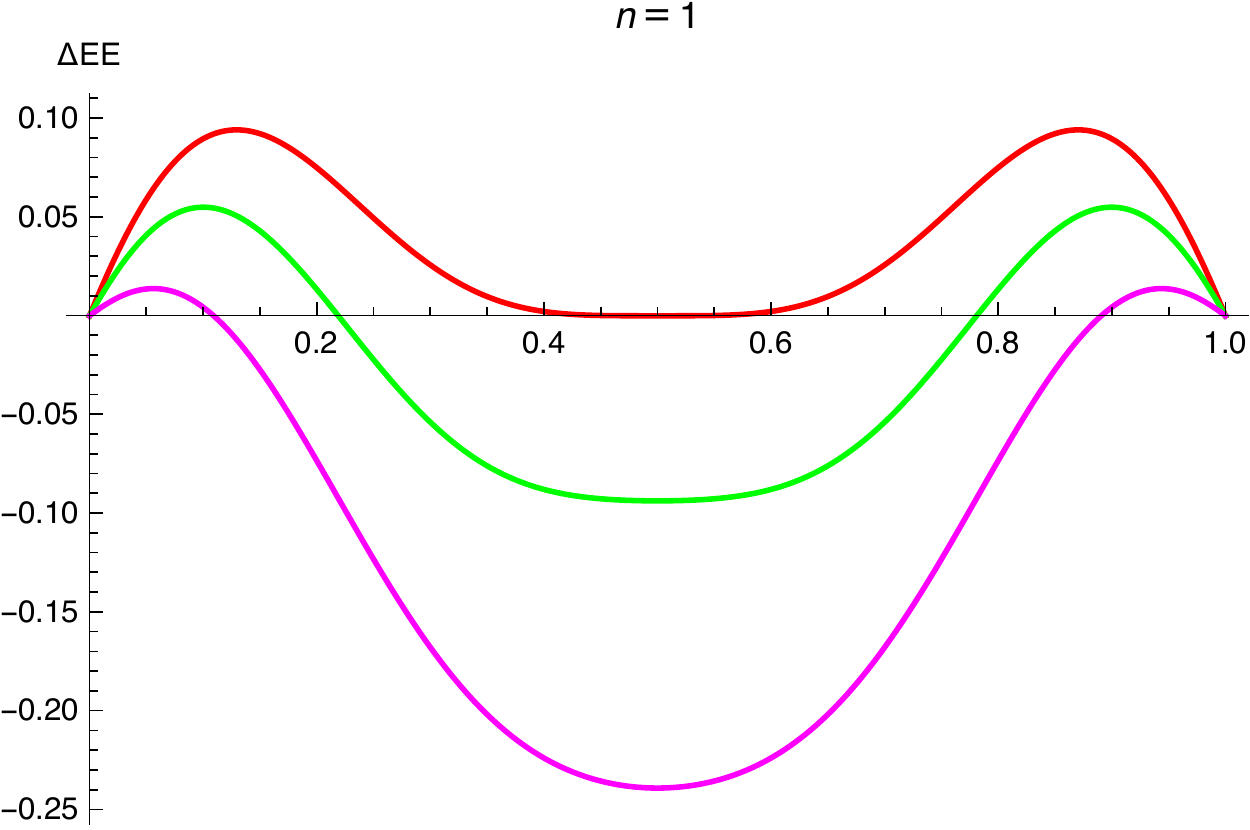}
    \includegraphics[width=7.5cm]{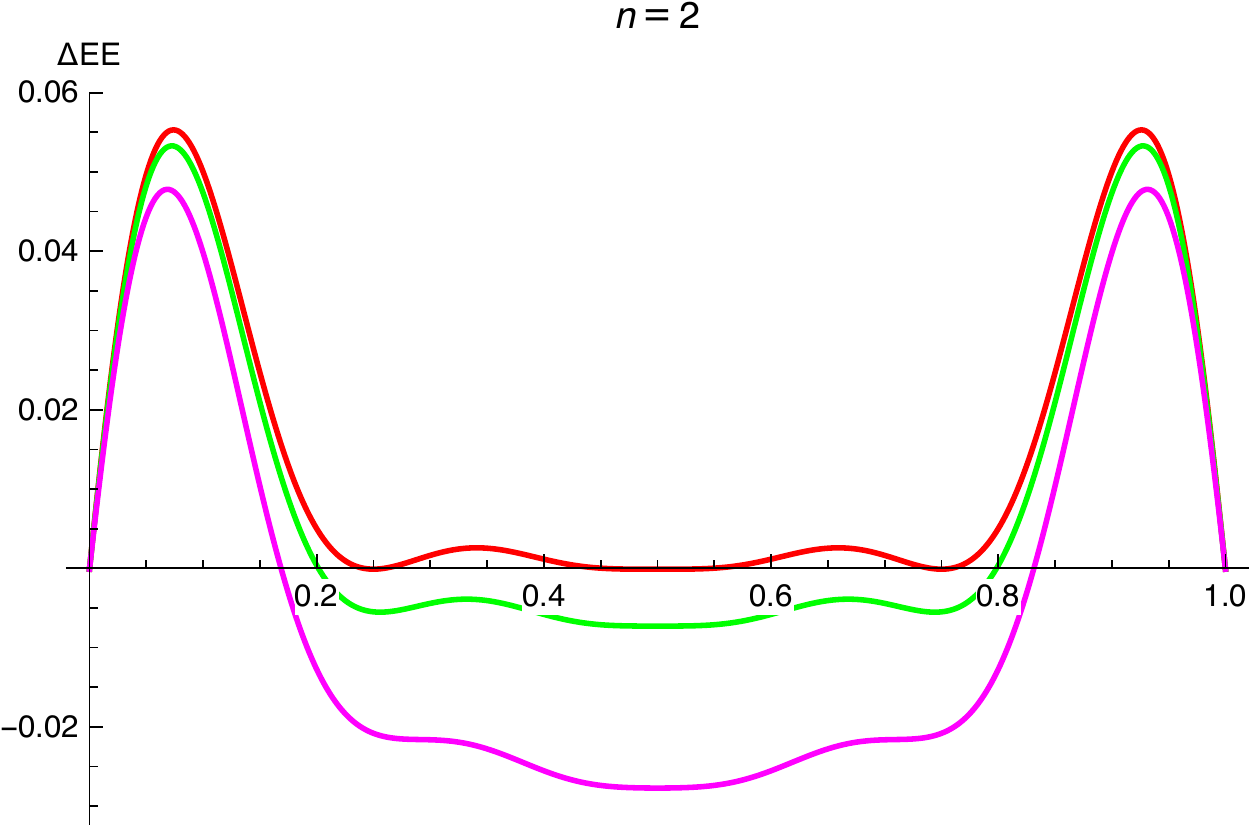}
       \caption{In this figure we present the difference between the EE in the BMN limit and twice the 
       EE of the single magnon \eqref{magnonEE}, 
       when we increase the mode number (from mode number one to two), for different values of $\lambda'$. 
       For the left plot the correspondence colour/$\lambda'$ is 
       Red $\Rightarrow$ $\lambda'=0$, Green $\Rightarrow$ $\lambda'=0.1$  \& Magenta $\Rightarrow$ $\lambda'=0.5$, 
       while for the right plot is 
       Red $\Rightarrow$ $\lambda'=0$, 
       Green $\Rightarrow$ $\lambda'=0.01$ \& Magenta $\Rightarrow$ $\lambda'=0.05$.}
   \label{fig:2.2}
\end{figure}

As we pointed out in \eqref{upperbound} and verified with the calculations of the EE for the three different rank one sectors 
in section \ref{EE-2magnons}, the EE of a single magnon multiplied by the number of impurities (two in our case) appears  
to be an upper bound for the EE. As can be seen from the two plots of figure \ref{fig:2.2} this bound is violated as soon as 
we move away from the $\lambda'\rightarrow 0$ limit. In order to underline the violation of this bound for finite $\lambda'$
in figure \ref{fig:2.2} we plot the the difference between the EE in the BMN limit and twice the EE of the single magnon, 
when we increase the mode number (from mode number one/left to two/right), for different values of $\lambda'$. 
As can be seen from the plots, when the coupling increases it is possible to find pieces of the spin chain with 
EE that violate the bound and as we increase the value of $\lambda'$ more and more pieces acquire EE that 
violate the bound. Gradually all the pieces of the spin chain will violate the bound and the 
bigger is the mode number the faster (i.e. with lower value of $\lambda'$) this violation will be implemented. 
The smaller the part of the chain we cut the more we have to increase the value of 
$\lambda'$ to violate the bound. In order to decide if it exists a $\lambda'_{crit}$, 
after which the bound is violated no matter how small is the length of the chain, we need to consider 
$1/J$ corrections to the EE in \eqref{EE-2mag-exact}.

\begin{figure}[h] 
   \centering
   \includegraphics[width=7.5cm]{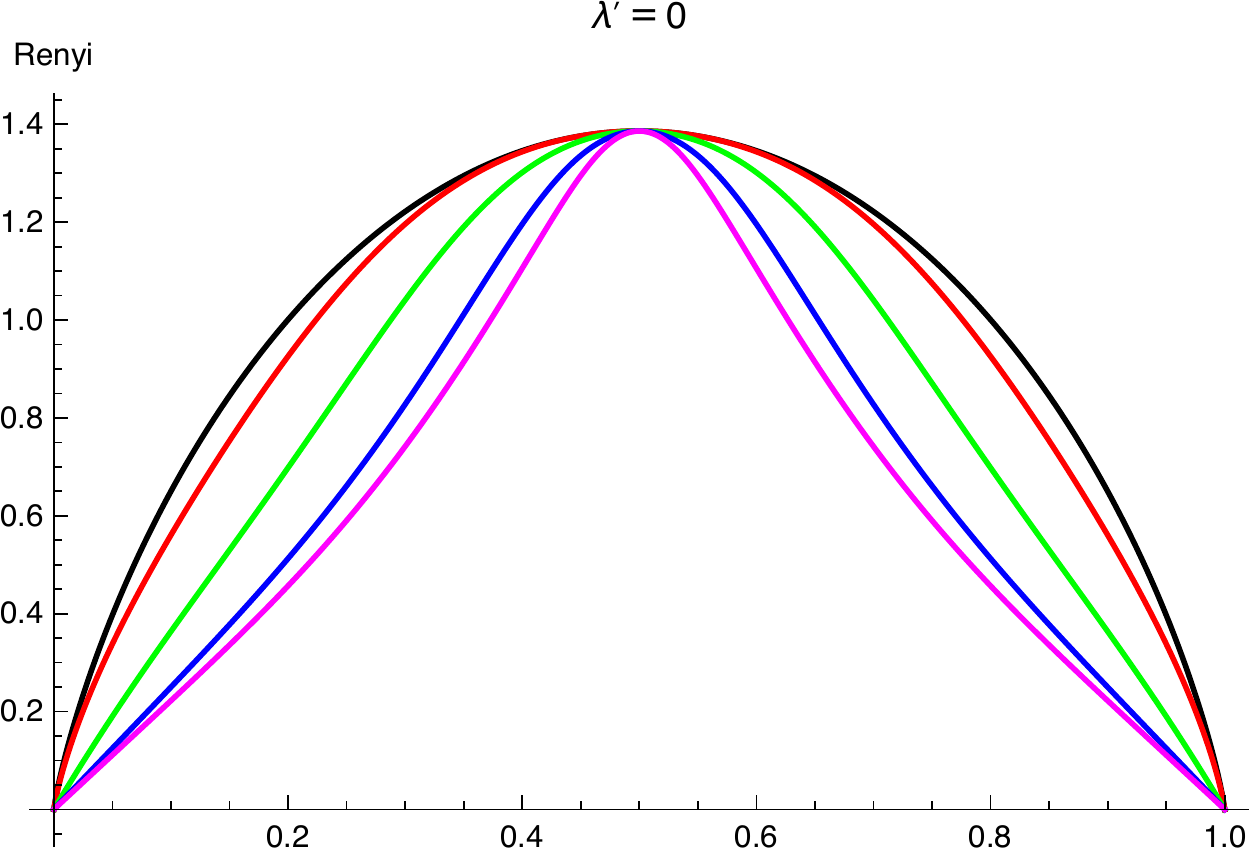}
    \includegraphics[width=7.5cm]{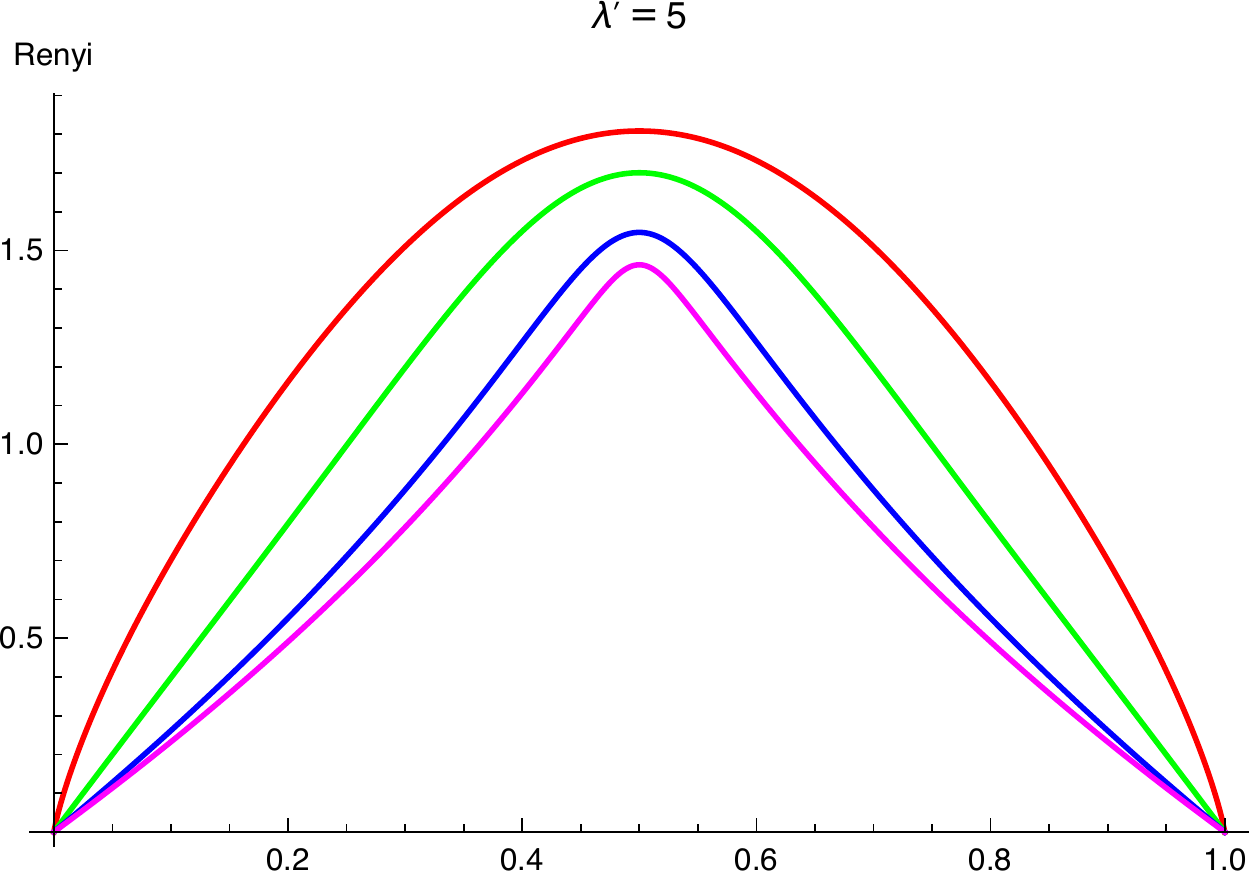}
       \caption{In this figure we present the Renyi entropy in the BMN limit (of different orders) 
       as a function of the normalised splitting,
       for two values of $\lambda'$, $\lambda'=0$ and $\lambda'=5$. The correspondence between color and 
       Renyi order is the following:  Red $\Rightarrow$ $\eta=1$, Green $\Rightarrow$ $\eta=2$,  
       Blue $\Rightarrow$ $\eta=5$ \& Magenta $\Rightarrow$ $\eta=10$. The Black curve corresponds to 
       twice the EE of the single magnon. The calculations are for mode number $n=1$.}
   \label{fig:2.3}
\end{figure}

We close this section with a couple of plots for the Renyi entropy and its dependence on the order $\eta$, 
$\lambda'$ and the normalised splitting $N/L$. In figure \ref{fig:2.3} 
we present the Renyi entropy (of different orders) as a function of the normalised splitting,
for two values of $\lambda'$, $\lambda'=0$ and $\lambda'=5$. From these two plots it is clear that for finite $\lambda'$ 
the bound is violated for any value of the order parameter $\eta$. Furthermore for two order parameters $\eta_1$ \& $\eta_2$ 
with $\eta_1<\eta_2$ the two Renyi entropies  $S_R^{(\eta_1)}$ \& $S_R^{(\eta_2)}$ obey the inequality  
$S_R^{(\eta_1)}>S_R^{(\eta_2)}$, making $S_R^{(2)}$ a useful lower bound on $S_R^{(1)}$. 
This is a known feature of the Renyi entropy from field theory considerations of its functional dependence 
on the order parameter, (see e.g. \cite{Beck}).

\begin{figure}[h] 
   \centering
   \includegraphics[width=7.5cm]{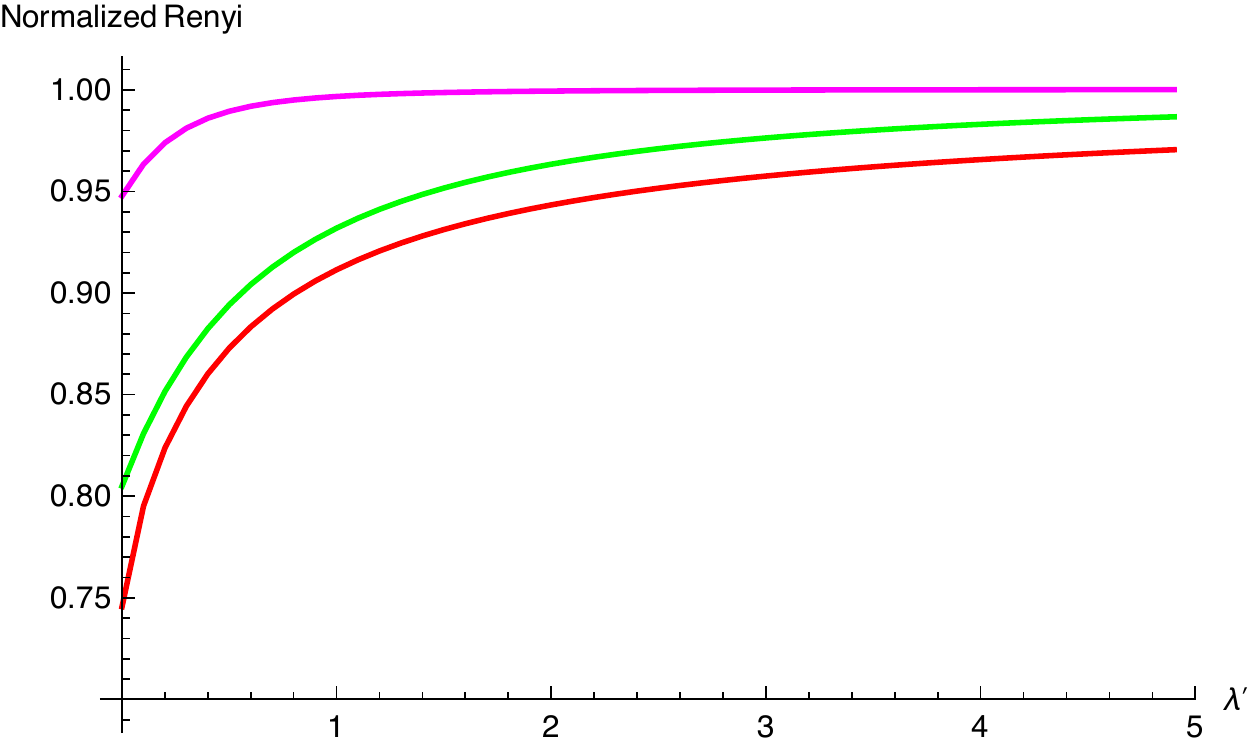}
    \includegraphics[width=7.5cm]{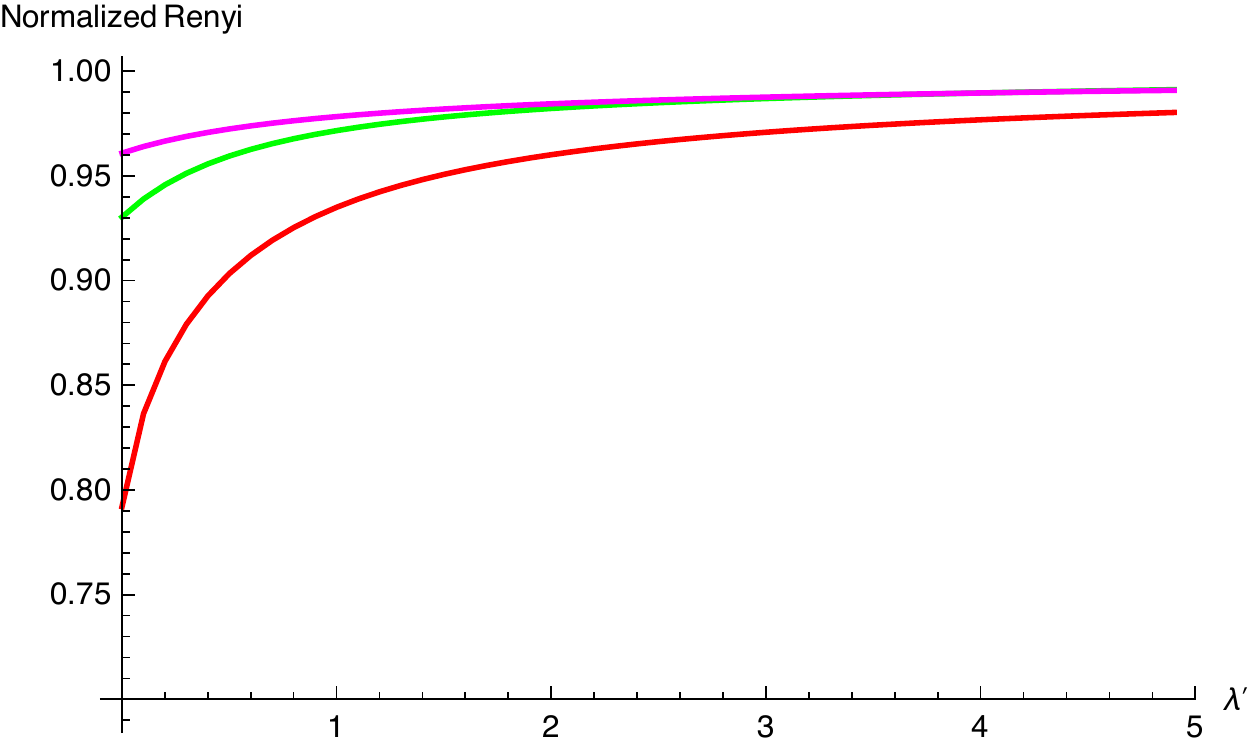}
       \caption{In this figure we present the Renyi entropy flow (of different orders) from the IR (lower values of $\lambda'$) 
       to the UV (higher values of $\lambda'$), as we change the normalised splitting (left plot $\Rightarrow$ $N/L=0.5$ 
       \& right plot $\Rightarrow$ $N/L=0.2$). As in figure \ref{fig:2.1}, we 
       have normalized each of the curves by dividing with the Renyi entropy for $\lambda'\rightarrow \infty$. 
       The correspondence between color and Renyi order is the following: Red $\Rightarrow$ $\eta=1$, 
       Green $\Rightarrow$ $\eta=2$ \& Magenta $\Rightarrow$ $\eta=10$.}
   \label{fig:2.4}
\end{figure}

In figure \ref{fig:2.4} we highlight the flow of the Renyi entropy from the UV to IR 
(besides that of the EE that we already saw in figure \ref{fig:2.1}). In this figure 
we present the normalised Renyi entropy (see \eqref{normal-def} for the definition of this normalisation) flow 
(of different orders) from the IR (lower values of $\lambda'$) to the UV (higher values of $\lambda'$), 
as we change the normalised splitting. From these plots it is clear that increasing the order 
(or decreasing the length of the spin chain we cut) decreases the difference between the IR 
and the UV values of the Renyi entropy.


\section{Entanglement Entropy from the Algebraic Bethe Ansatz}
\label{ABA-gen}

In this section we concentrate on the case where there is an arbitrary number of excitations (magnons) 
propagating in the spin chain. In such a case the method used in the previous section for the case of two magnons 
becomes cumbersome because of the large number  of excitations.   However, it is the formalism of 
ABA that comes to rescue in this occasion.  
Following the same reasoning as before, we split the spin chain of length $L$ in two pieces. 
The first piece which we call $D$ contains the sites from $1$ to $N$, while its complementary $D^C$
contains the sites from $N+1$ to $L$. In the  ABA language the wavefunction $|\psi\rangle$ describing the 
excited state is characterised by  a set of $M$ numbers $\{u_i\},\,\, i=1,2,...,M$. 
The $u_i$'s are called the rapidities of the $M$ magnons and for an on-shell state they should satisfy 
the Bethe equations of the  corresponding sector. Thus (see \cite{Faddeev:1996iy, Escobedo:2010xs} for all the details about the formalism of the ABA),
\begin{equation}\label{wave-ABA}
|\psi\rangle \, = \, |\{u_i\}\rangle \, = \, \sum_{a \bigcup \bar{a} \, = 
\, \{u_i\}} H(a, \bar{a})\, |a_l\rangle \otimes |\bar{a}_r\rangle \, , 
\end{equation} 
where the sum is over all the possible partitions of the $M$ magnons into two sets. 
If for example we have two magnons, as in the cases of the CBA we worked so far, the possible partitions are 
$\left( \{\},\{u_1,u_2\} \right) ,  \left( \{u_1\},\{u_2\} \right),  \left( \{u_2\},\{u_1\} \right)$ \&  $\left( \{u_1,u_2\}, \{\} \right)$.
The set of magnons with rapidities $a_l$  are sitting in the left part of the spin chain, that is in region $D$,  
while the set of magnons with rapidities $a_r$ are sitting in the complementary region of the chain, that is $D^C$. 
The function $H(a, \bar{a})$ describes the weight of each partition, it is different in each rank one subsector 
of the ${\cal N}=4$ SYM and its form is given by (for the various functions appearing in \eqref{Hdef1} see also Appendix \ref{app:ABA})
\begin{equation} \label{Hdef1}
H(a,\bar{a}) \, = \, f^{a \bar{a}} \, d_r^a \, a_l^{\bar{a}}\, .
\end{equation}

Starting from \eqref{wave-ABA} it is now straightforward to evaluate the RDM
\begin{eqnarray} \label{RDM-ABA}
\rho_D \, &=& \, \kappa \, \sum_{\bar{c}_r} \,  
\sum_{\scriptsize \begin{array}{c} a_l\cup \bar{a}_r \\ b_l\cup \bar{b}_r \end{array}} \,
H(a_l, \bar{a}_r) \, H^*(b_l, \bar{b}_r) \, 
\langle \bar{c}_r| \bar{a}_r\rangle \otimes |a_l \rangle
\langle b_l | \otimes \langle \bar{b}_r |\bar{c}_r \rangle
\nonumber \\
&=& \kappa \, \sum_{m=0}^M \, \sum_{a_l^m \, , \, b^m_l} f^{(m)}_{a^m_l b^m_l}  \,\,
|a^m_l \rangle \langle b^m_l | \, , 
\end{eqnarray}
where we have defined the following quantity
\begin{equation}
f^{(m)}_{a^m_l b^m_l} \, \equiv \, H(a^{(m)}_l, \bar{a}^{(M-m)}_r) \, H^*(b^{(m)}_l, \bar{b}^{(M-m)}_r) \, \, 
\langle \bar{b}^{(M-m)}_r | \bar{a}^{(M-m)}_r \rangle.
\end{equation} 
In \eqref{RDM-ABA} $\bar{c}_r$ denotes a complete basis of states of the complementary part of the spin chain $D^C$. 
In order to get  the last expression for the RDM we have used the completeness relation for the basis $\bar{c}_r$
\begin{equation}
\sum_{\bar{c}_r} \, \, |\bar{c}_r \rangle\langle \bar{c}_r| \, = \,  I_r\, . 
\end{equation}
Here we should stress that because the scalar product of two states involving different numbers of magnons is zero 
the RDM can be written as a sum of terms each of which has a definite number of excitations $m$ in the region $D$. 
Subsequently, for each of these terms one has to sum over all possible partitions of the full set of rapidities into two sets, 
one having $m$ excitations and its complementary having $M-m$ excitations. 
Notice that the scalar products appearing in the second line of \eqref{RDM-ABA} are generic off-sell products, 
since none of the sets of rapidities $\bar{a}_r$ nor $\bar{b}_r$ satisfy the Bethe equations for the complementary 
region of the spin chain. This scalar product is given in terms of the recursion relation in equation \eqref{rec_relation}.
The normalisation constant $\kappa$ is obtained by demanding the condition
\begin{eqnarray} \label{normal-ABA}
&&Tr_D\rho_D \, = \, 1 \qquad \Rightarrow \qquad 
\kappa^{-1} \, = \, \langle\psi|\psi\rangle \, = \, \sum_{m=0}^M \, \sum_{a_l^m}F^{(m)}_{a^{m}_l a^{m}_l} \, , 
\nonumber\\
&&\text{with} \quad F^{(m)}_{a^{m}_l b^{m}_l} \, = \, f^{(m)}_{a^{m}_l c^{m}_l} \, g^{(m)}_{c^{m}_l b^{m}_l}
\qquad {\rm and} \qquad 
g^{(m)}_{c^{m}_l b^{m}_l} \, = \, \langle c^{(m)}_l| b^{(m)}_l\rangle \, . 
\end{eqnarray} 
A final comment concerns the dimensionality of the matrices $F^{(m)}$, $f^{(m)}$ and $g^{(m)}$. 
The dimensionality of each of these square matrices depend on the number $m$ of magnons sitting 
in the region $D$ and is given by 
\begin{equation}
d(m) \, \equiv \, \frac{M!}{m!(M-m)!} \, .
\end{equation}
As we will see in a while it is the matrices $F^{(m)}$ which one has to diagonalise in order to calculate the EE. 
As a result the numerical complexity of the calculation grows not with the number of the sites of the region $D$ 
of which the EE we are after, but like $2^M$, that is with the number of magnons running in the spin chain. 
This advantage is related to the fact that we have employed the powerful technique of the ABA and not that of 
the CBA.

We are now in position to evaluate the $\eta$-th power of the RDM and take its trace to obtain the Renyi 
Entropy of the region $D$ as follows
\begin{equation}
\rho^{\eta}_D \, = \, \kappa^{\eta} \, \sum_{m=0}^M \left[\rho^{(m)}_D \right]^n 
\quad {\rm with} \quad 
\left[\rho^{(m)}_D\right]^{\eta} \, = \, \sum_{a^{m}_l, b^{m}_l,c^{m}_l } |a^{(m)}_l \rangle \langle b^{(m)}_l |  
\left[F^{(m)}_{a^{m}_l c^{m}_l}\right]^{\eta-1}f^{(m)}_{c^{m}_l b^{m}_l} \Rightarrow
\nonumber
\end{equation}
\begin{equation} \label{rton-ABA}
Tr_D \left[\rho_D\right]^{\eta} \, = \, \kappa^{\eta} \, \sum_{m=0}^M \sum_{a^m_l} \, 
\left[ F^{(m)}_{a^m_l a^m_l} \right]^{\eta} \, . 
\end{equation} 
Thus we see that in order to evaluate the Renyi entropy
we need to diagonalise the matrices $\kappa \, F^{(m)}$ and sum their eigenvalues 
after they are raised to the  $\eta^{th}$ power.
By taking the $\eta \rightarrow 1$ limit of the Renyi entropy it is straightforward to show that the EE is given by
\begin{equation} \label{EE-ABA}
S_{EE} \, =  
\, - \,  \sum_{m=0}^M \, Tr\Big[\kappa \, F^{(m)} \, \log{(\kappa \, F^{(m)})}\Big] \, .
\end{equation}
Thus, as mentioned above, it is enough to diagonalise each of the matrices  $\kappa F^{(m)}$. 
If we denote the eigenvalues of each of these matrices by $\lambda^{(m)}_i$ with $i=1,2,\ldots,d(m)$ then the EE
of the part $D$ of the spin chain can be finally written as\footnote{To be precise we should notice the following: the upper limit in the second summation in 
\eqref{EE-ABA-fin} is strictly speaking $\min\{M, N \}$ and not $M$. Similarly, the lower limit is  $\max\{0, M-(L-N)\}$. 
The reason is that if $N$  is less than $M$ then in the calculation of the EE contribute only a portion of the 
possible partitions of the Bethe roots and not the whole.}
\begin{eqnarray}\label{EE-ABA-fin}
S_{EE} \, = \, - \, \sum_{m=0}^M \, \sum_{i=1}^{d(m)} \, \lambda^{(m)}_i  \, \log{\lambda^{(m)}_i} \, . 
\end{eqnarray} 

\begin{figure}[h] 
   \centering
   \includegraphics[width=7.5cm]{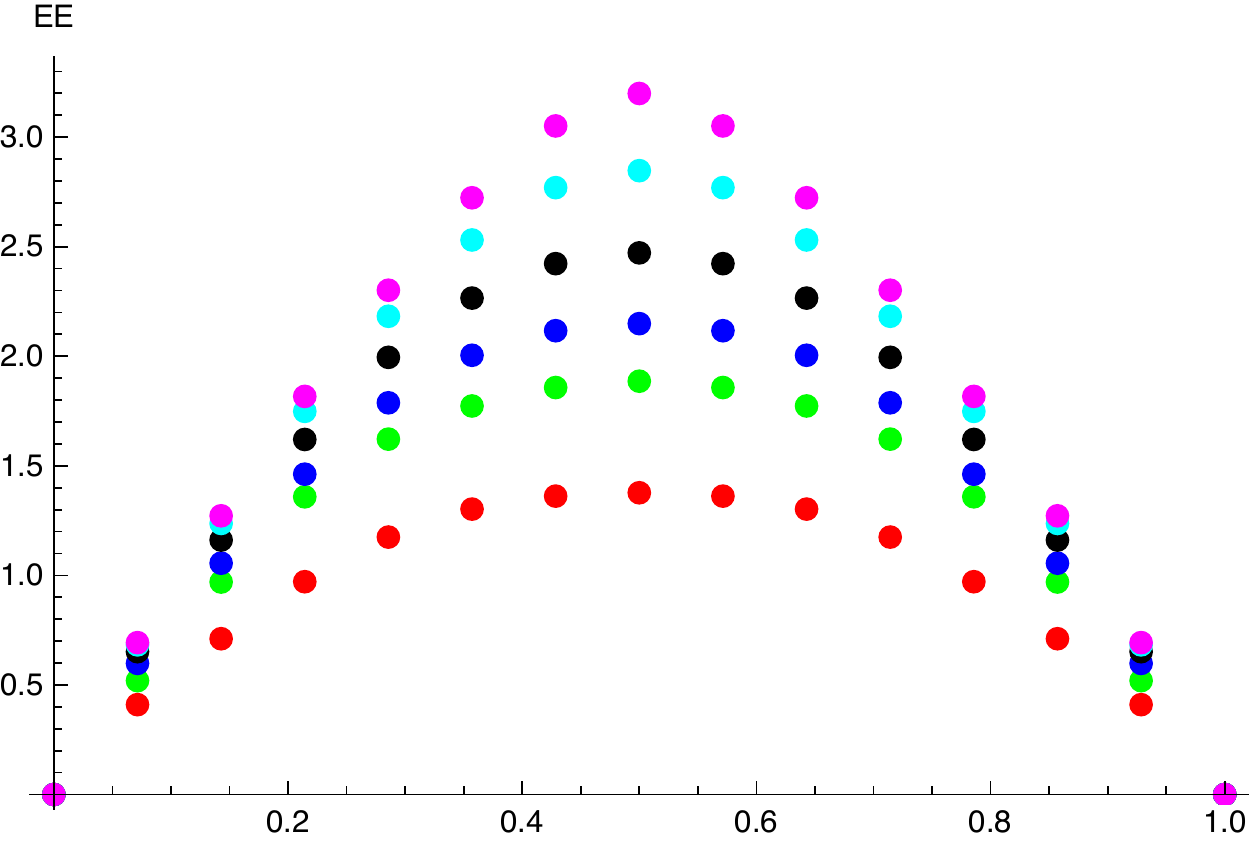}
    \includegraphics[width=7.5cm]{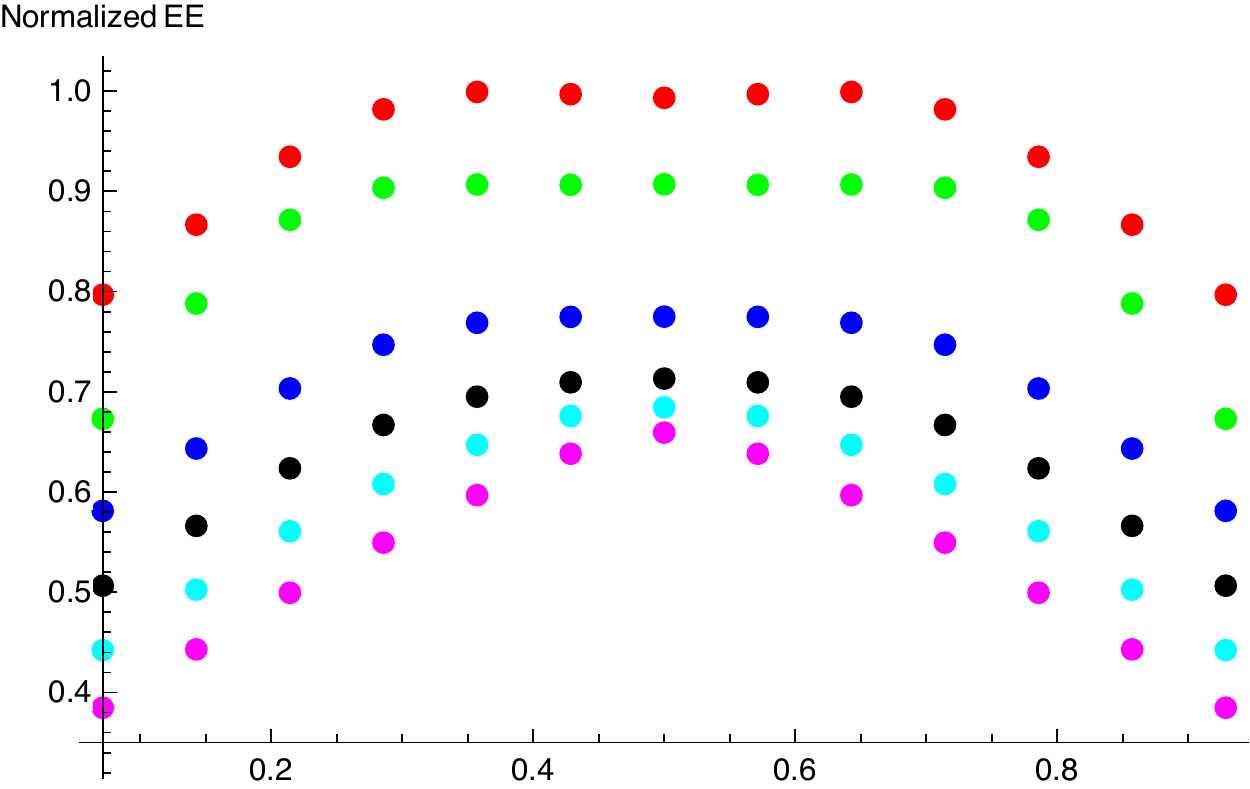}
       \caption{In this figure we present the EE of an excited state in the $SU(2)$ sector. On the left part of the figure 
       it is the plot of the EE as a function of the normalised splitting of the spin chain. 
       On the right part of the figure it is the plot of the normalised EE (i.e. the EE for $M$ magnons divided by $M$ times 
       the EE of one magnon) again as a function of the normalised splitting.
       The different colours for the bullets in both plots correspond to different number of magnons present in the spin chain. 
       The correspondence between colour and number of magnons is the following: Red $\Rightarrow$ 2 magnons, 
       Green $\Rightarrow$ 3 magnons, Blue $\Rightarrow$ 4 magnons, Black $\Rightarrow$ 5 magnons, 
       Cyan $\Rightarrow$ 6 magnons \& Magenta $\Rightarrow$ 7 magnons.}
   \label{fig:3.1}
\end{figure}

\begin{figure}[h] 
   \centering
   \includegraphics[width=7.5cm]{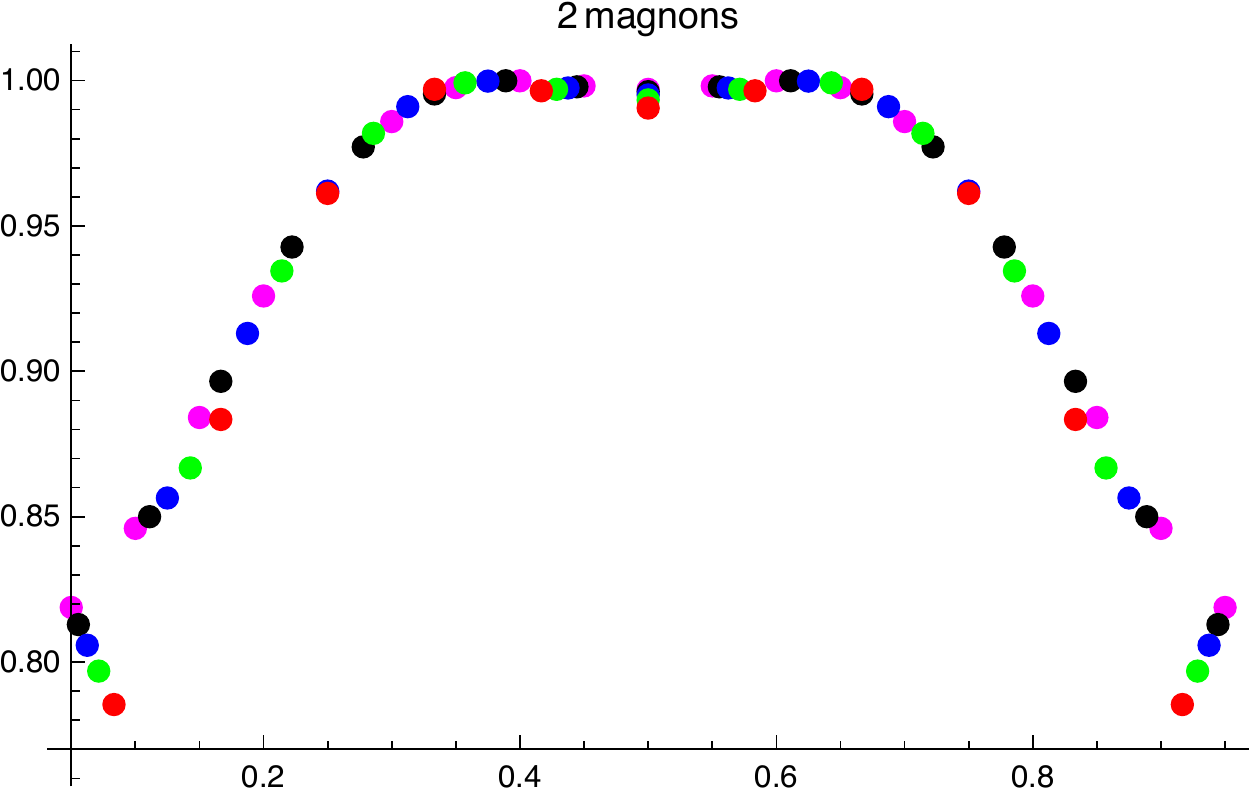}
    \includegraphics[width=7.5cm]{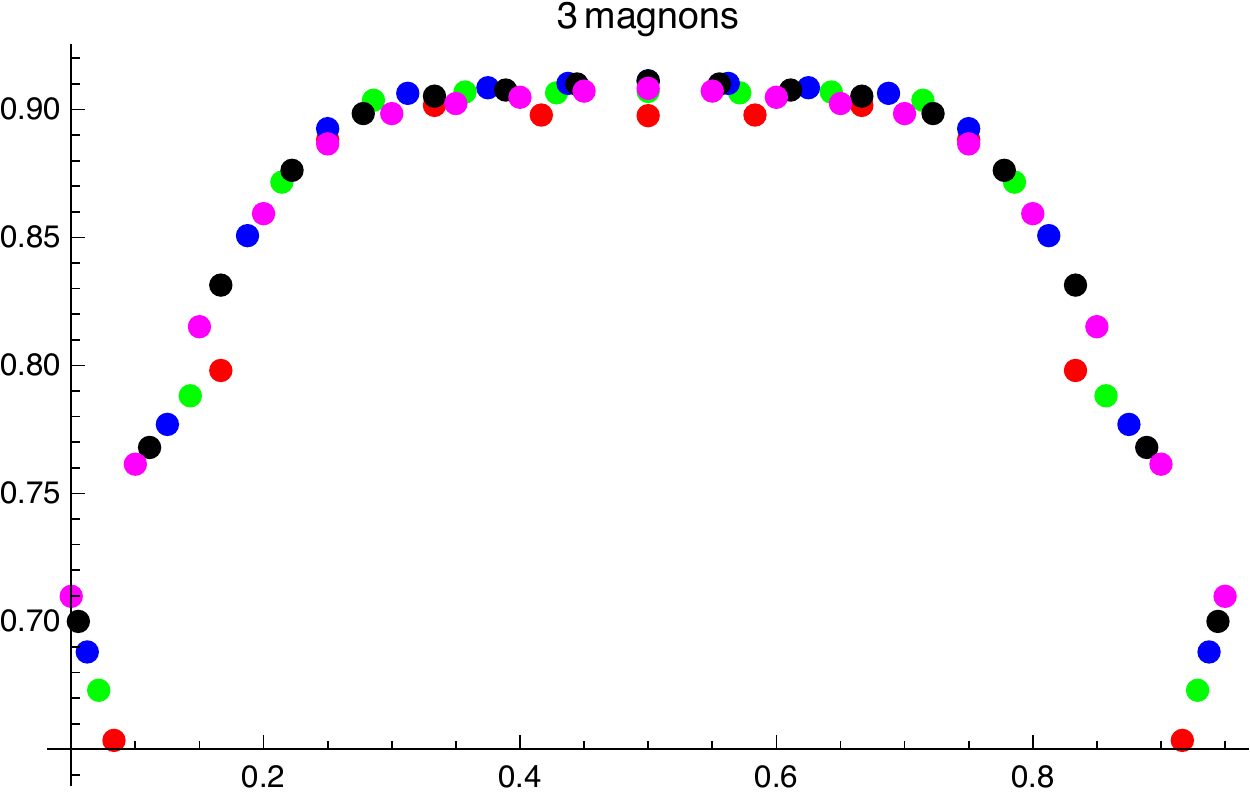}
    \includegraphics[width=7.5cm]{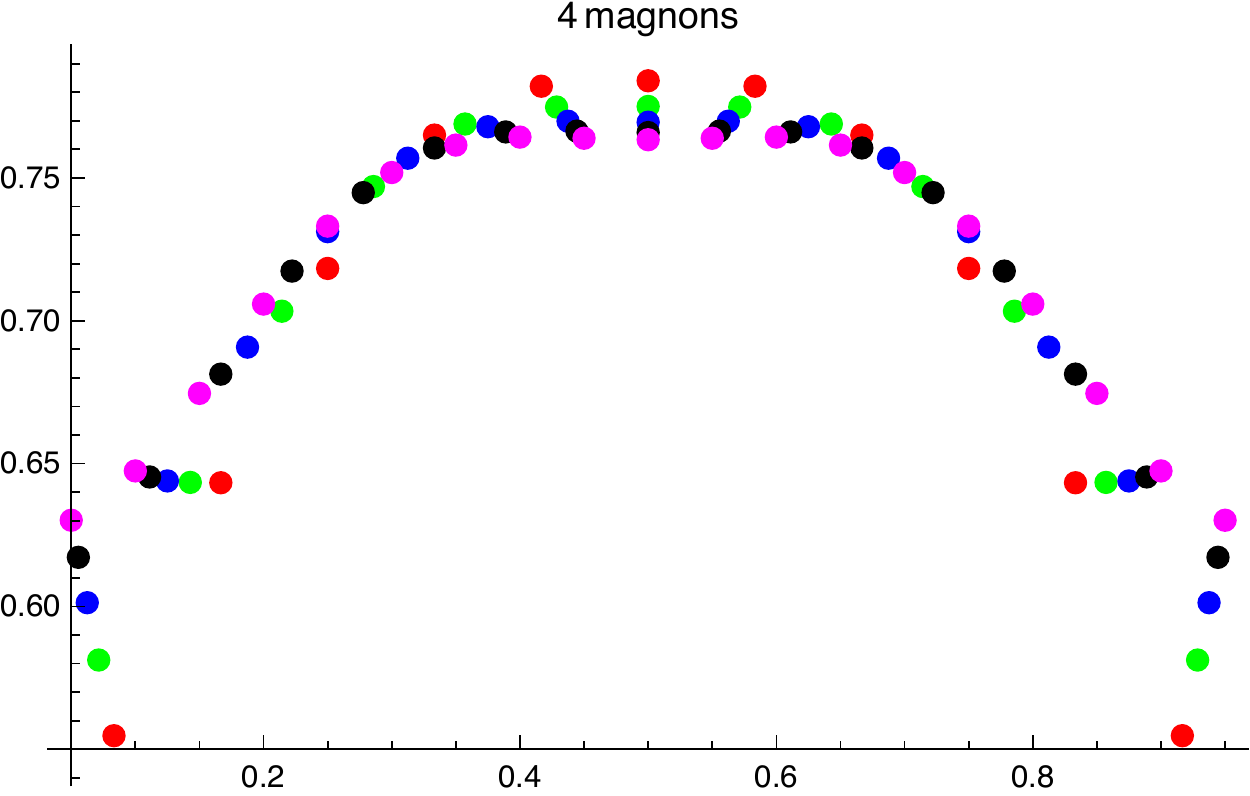}
      \includegraphics[width=7.5cm]{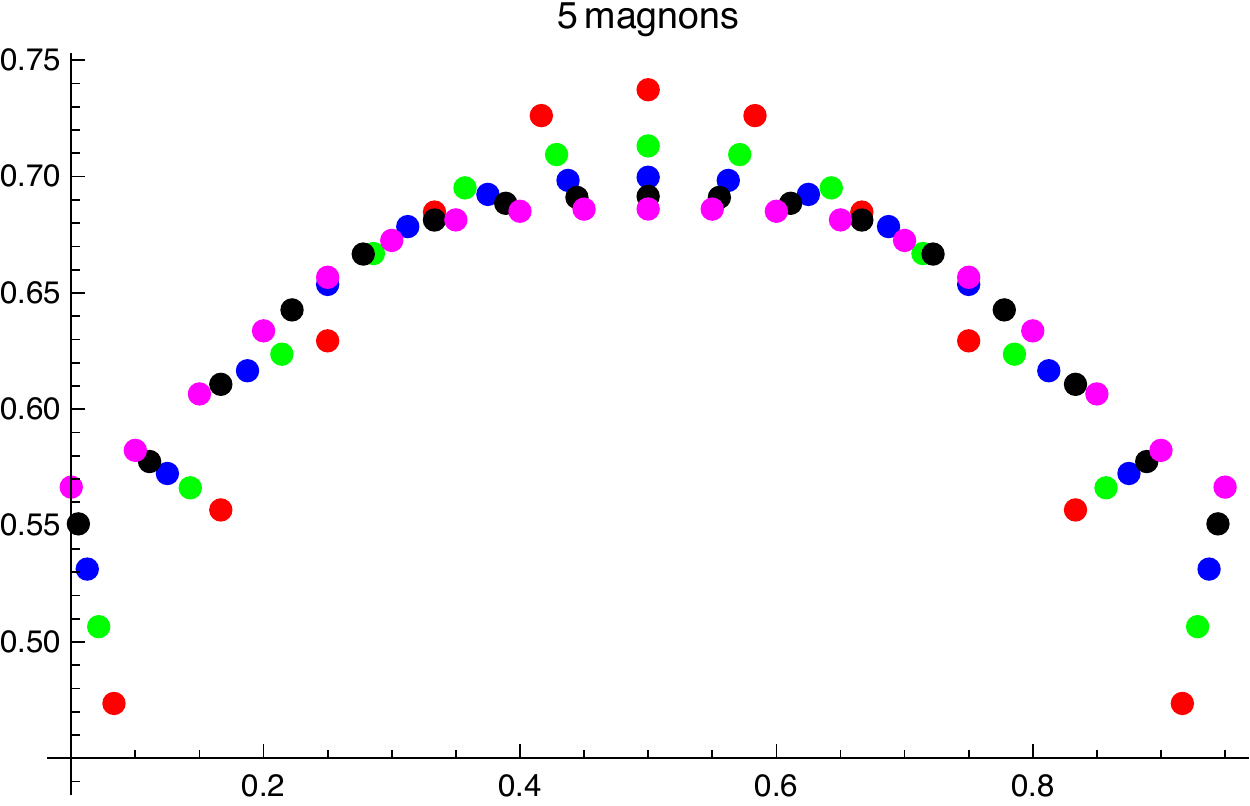}
    \includegraphics[width=7.5cm]{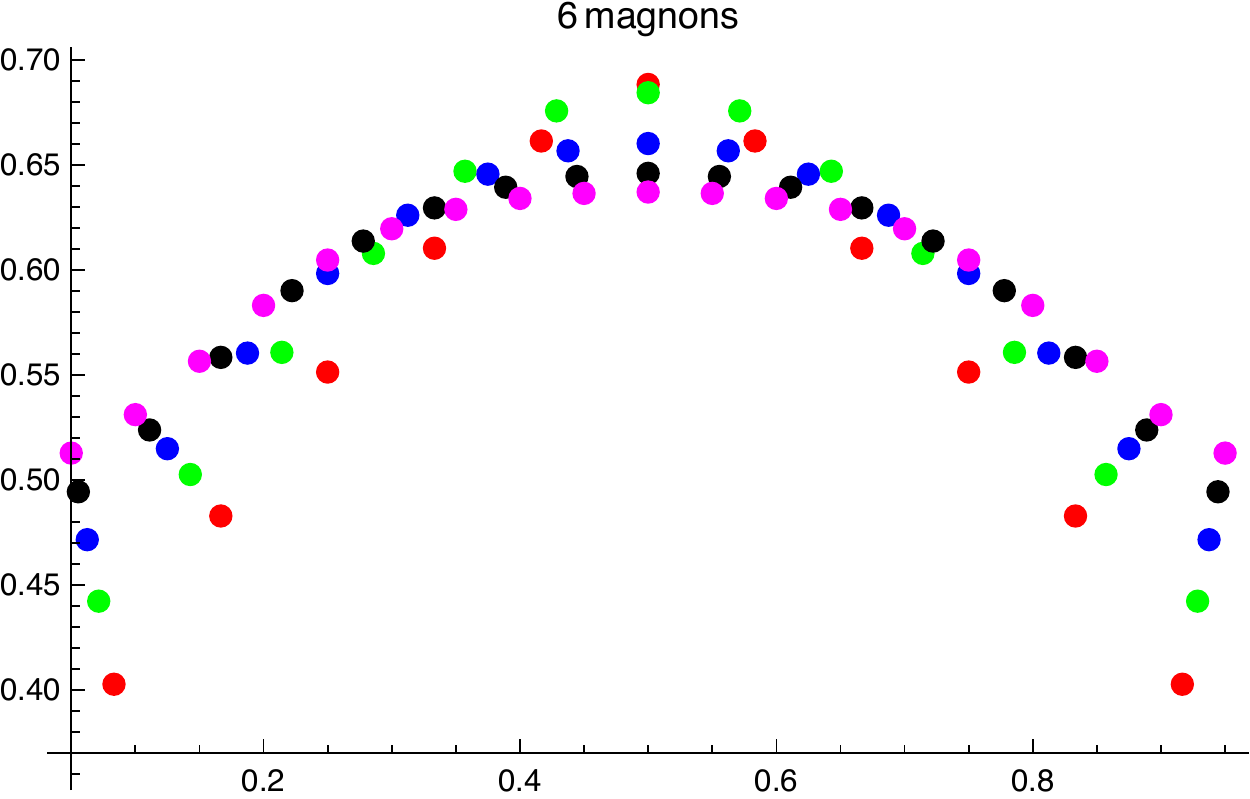}
    \includegraphics[width=7.5cm]{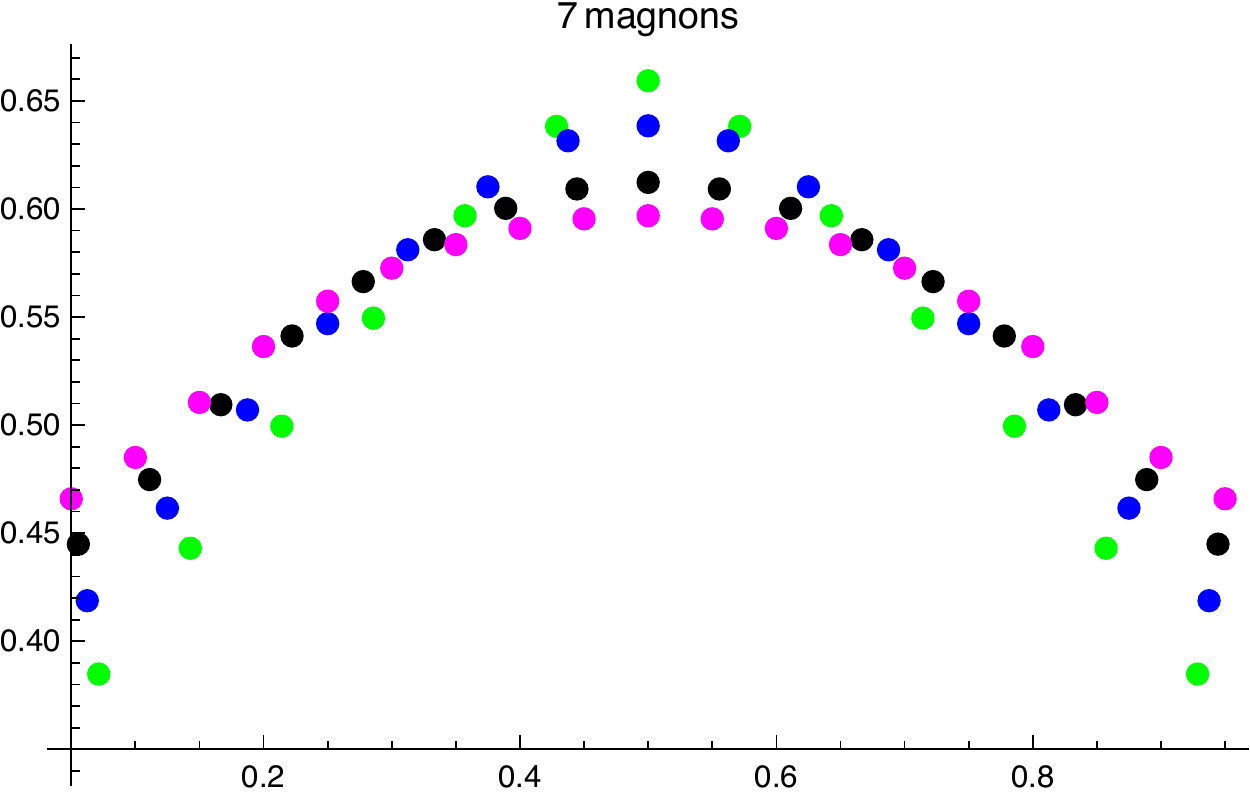}
       \caption{In this figure we present the normalised EE for different number of magnons when the length of 
       the spin chain changes from $L=12$ to $L=20$.  The correspondence between colour and 
       length of the spin chain is the following: Red $\Rightarrow$ $L=12$, 
       Green $\Rightarrow$ $L=14$, Blue $\Rightarrow$ $L=16$, Black $\Rightarrow$ $L=18$
       \& Magenta $\Rightarrow$ $L=20$.}
   \label{fig:3.2}
\end{figure}

In the final expression \eqref{EE-ABA-fin}, the EE of each one of the three rank one sectors of ${\cal N} =4$ SYM 
is a function of the length of the spin chain, the position we split it in two parts 
($D$ and its complement  $D^C$) and the number of excitations (magnons). 
In figures \ref{fig:3.1} \& \ref{fig:3.2} we focus on the $SU(2)$ sector of the theory.\footnote{The rapidities of the excited states, 
for which we calculate the EE, are listed in Appendix \ref{app:BR}.} 
On the left part of figure \ref{fig:3.1} 
we plot the EE of a spin chain with 14 sites as a function of the position of the splitting point.
We plot the EE for different number of magnons, from two to seven. 
At this point we should point out that there is a perfect match of the EE for the two magnons,
between the CBA and the ABA calculations (for all the three rank-one subsectors). That is a non-trivial consistency 
check for the calculations that we present.  

The Bethe roots we use in the calculation of the EE are coming after solving the Bethe equation, with the recursive method 
of \cite{Bargheer:2008kj}, and distribute themselves along two disjoint cuts on the complex plane\footnote{The 
MATHEMATICA code to explicitly produce those solutions can be found in Appendix E1 of \cite{Escobedo:2012ama}.}. 
As can be seen from this plot the EE increases as we increase the number of the magnons, 
but the qualitative behaviour remains the same.  It increases as the splitting point approaches the middle 
of the chain and it is symmetric under the change $N$ (splitting point) $\rightarrow$ $L-N$. 

Our initial motivation in employing the very efficient ABA formulation was reaching the thermodynamic limit, by 
increasing both the number of magnons and the length of the spin chain. However, this is a very complicated numerical task.  
As we mentioned before, the calculation of the EE for $M$ magnons in a spin chain 
boils down to diagonalising  matrices with dimension $d(m)$. This means that either a very powerful machine 
is needed or the problem needs to be formulated in a different basis when the number of magnons increases. 

On the right panel of figure \ref{fig:3.1} we plot the normalised EE (i.e. the EE for $M$ magnons divided by $M$ times 
the EE of one magnon) again as a function of the position of the splitting point. We plot the EE for different number 
of magnons and the behaviour is different to the one we noticed before. Now increasing the number of magnons 
decreases the normalised EE. 
Notice also in the same plot, that the EE for two magnons almost saturates the bound not in the 
middle of the spin chain but at the splitting points 5/14 and 9/14. The maximum of the curve of the normalised EE 
is not in the middle also for the three magnons, but  eventually as we increase their number 
this maximum moves to the centre.  

Until now we kept constant the length of the spin chain, modifying the number of the magnons. 
In figure \ref{fig:3.2} we change this perspective, keeping constant the number of the magnons while 
changing the length of the spin chain.  As can be seen from all the plots of figure \ref{fig:3.2}, 
where we present the normalised EE (this is the only quantity that make sense to compare) for 
different number of magnons (from two to seven) when the length of the spin chain changes, we observe the following pattern.
When the number of the magnons is small (from two to four) the curve almost does not change as 
we change the length of the spin chain. Increasing the number of the magnons, we notice two effects depending 
on the length of the part of the spin chain we cut: 
When the length of the cut piece is small the EE increases when we increase the length of the spin chain, while the 
opposite happens for the EE when the cut piece is close to the half of the spin chain.

\begin{figure}[h] 
   \centering
   \includegraphics[width=7.5cm]{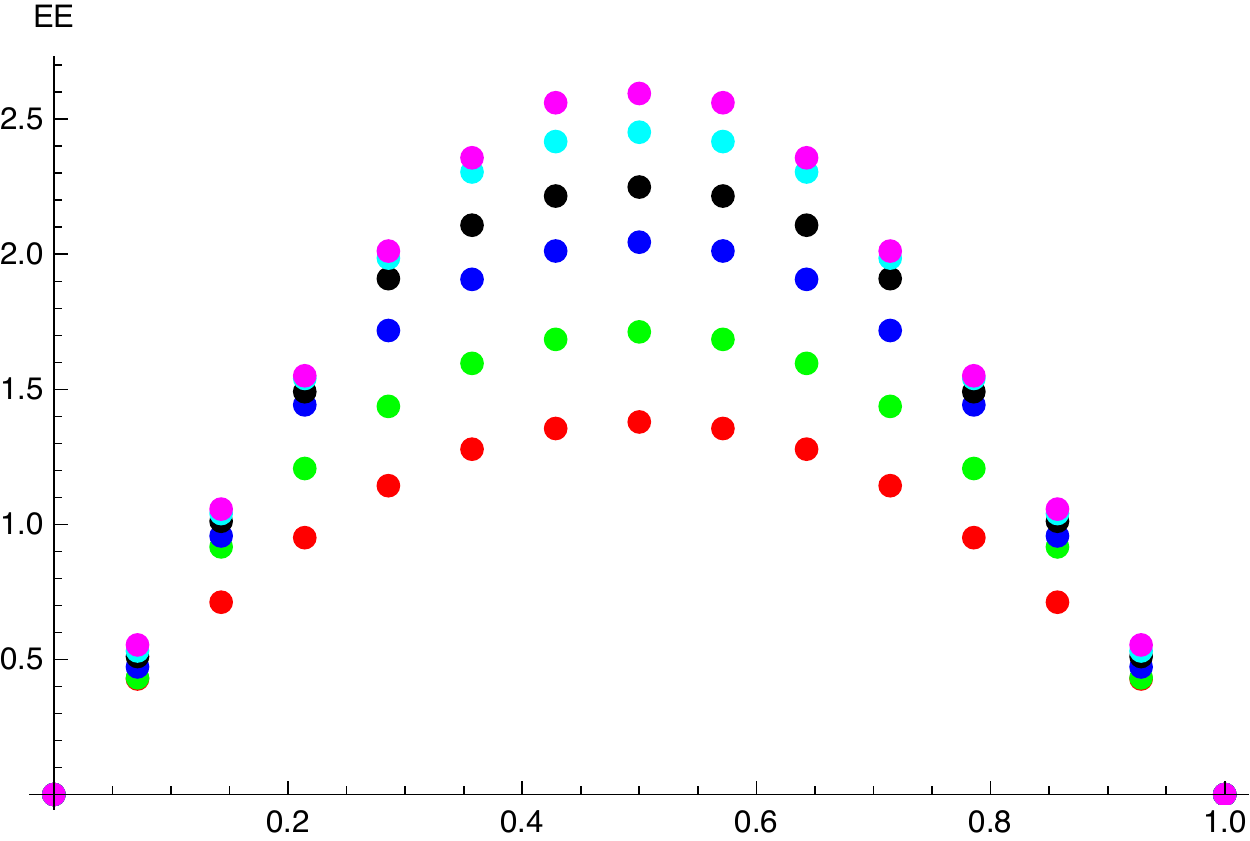}
    \includegraphics[width=7.5cm]{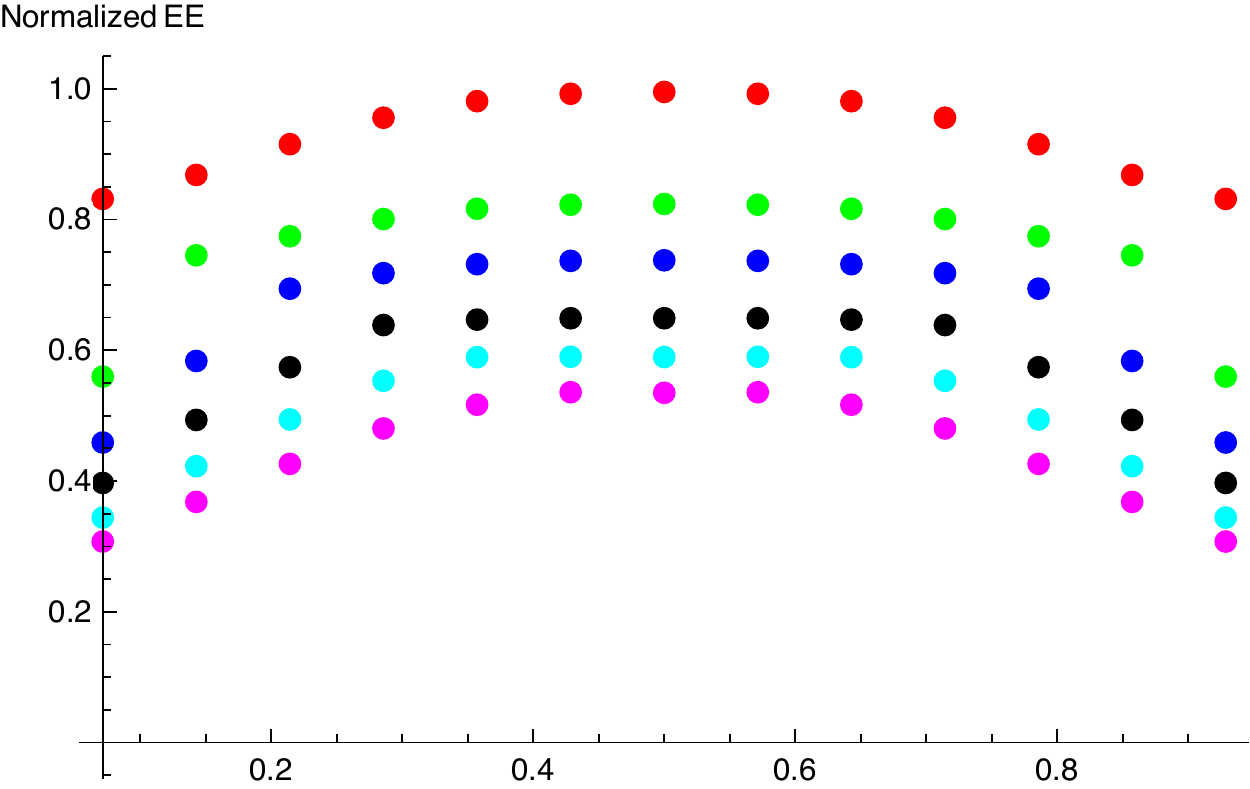}
       \caption{In this figure we repeat the calculations that we presented in figure \ref{fig:3.1}, 
       but for an excited state in the $SL(2)$ sector.  The Bethe roots are sitting in two cuts on the real axis.  The correspondence between colour and number of magnons is the following: Red $\Rightarrow$ 2 magnons, 
       Green $\Rightarrow$ 3 magnons, Blue $\Rightarrow$ 4 magnons, Black $\Rightarrow$ 5 magnons, 
       Cyan $\Rightarrow$ 6 magnons \& Magenta $\Rightarrow$ 7 magnons.}
   \label{fig:3.3}
\end{figure}

\begin{figure}[h] 
   \centering
   \includegraphics[width=7.5cm]{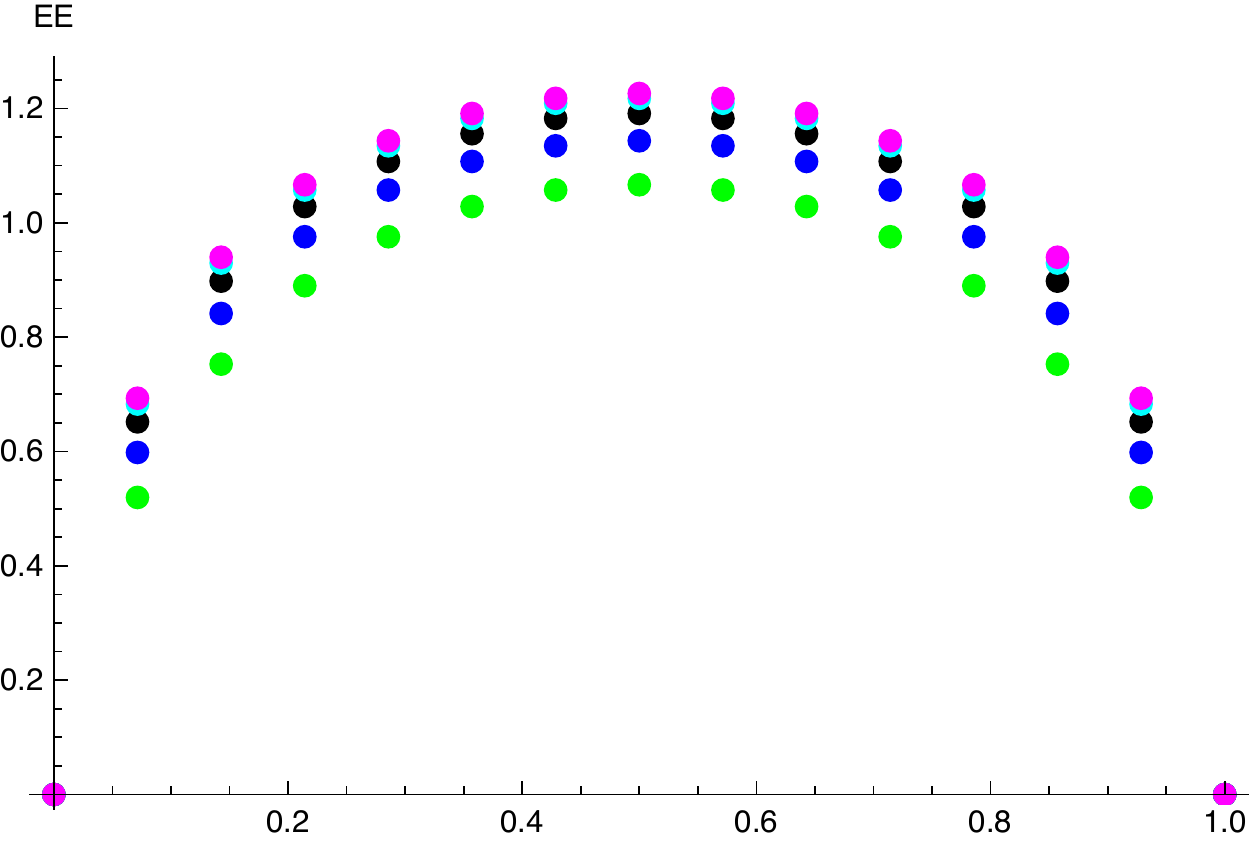}
    \includegraphics[width=7.5cm]{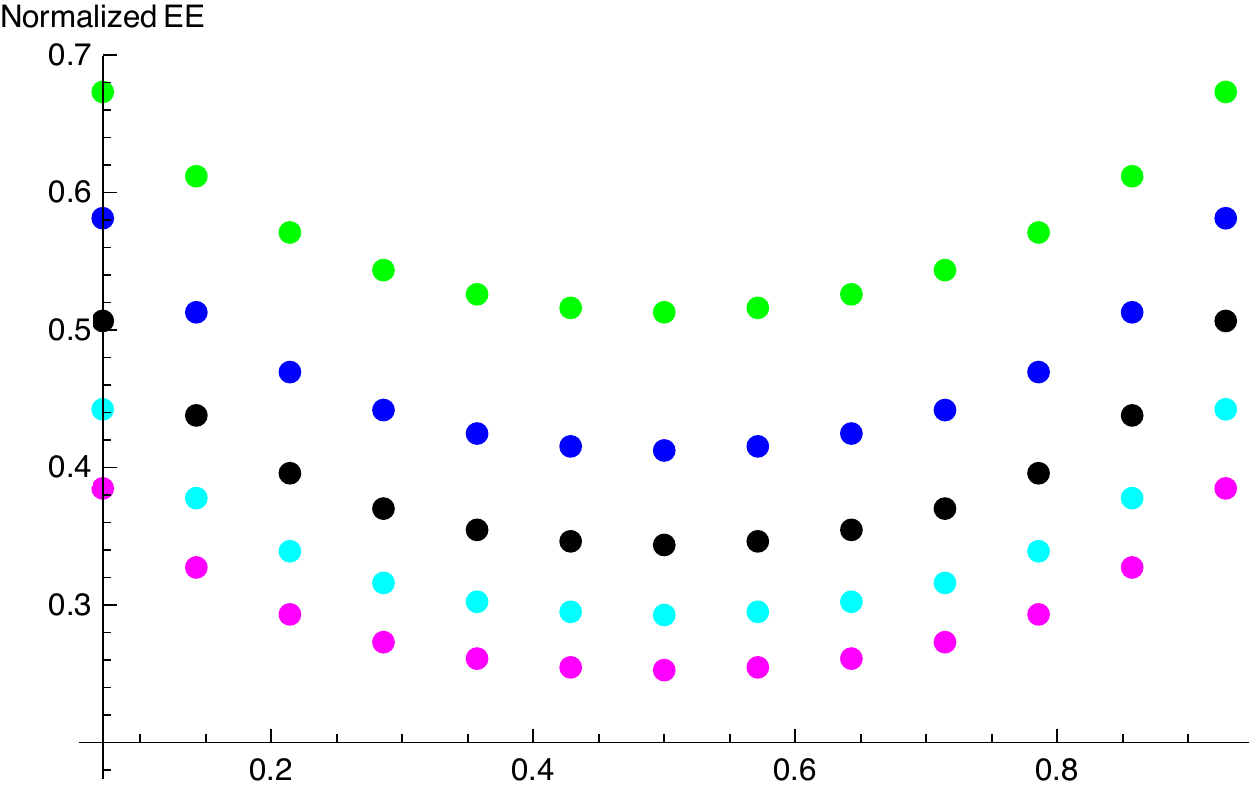}
       \caption{In this figure we repeat the calculations that we presented in figure \ref{fig:3.1}, 
       but for an excited state in the $SU(1|1)$ sector.  The correspondence between colour and number of magnons is the following: 
       Green $\Rightarrow$ 3 magnons, Blue $\Rightarrow$ 4 magnons, Black $\Rightarrow$ 5 magnons, 
       Cyan $\Rightarrow$ 6 magnons \& Magenta $\Rightarrow$ 7 magnons.}
   \label{fig:3.4}
\end{figure}

In figures \ref{fig:3.3} and \ref{fig:3.4} we complete the computation of the EE for the three rank one sectors 
of ${\cal N}=4$ SYM by repeating the calculation for the $SL(2)$ and the $SU(1|1)$ sectors, respectively.
In order to numerically perform these computations we use the same MATHEMATICA code as for the 
$SU(2)$ sector only changing accordingly the definitions for the functions $f$, $g$, $a$ \& $d$, as they appear 
in  equations \eqref{SL2def} \& \eqref{SU11def}, as well as the corresponding scattering matrices 
for the calculation of the Bethe roots.

The results for the EE of the $SL(2)$ sector are similar to the ones of the $SU(2)$ sector.
We should mention that we have chosen the excited states for which the Bethe roots are sitting 
in two symmetric cuts on the real axis.\footnote{The rapidities of the 
excited states,  are listed in Appendix \ref{app:BR}.} 
In the plots of the normalised EE there are though some differences. 
Notice that the EE for two magnons almost saturates the bound in the middle of the spin chain and not in some other points as 
in the $SU(2)$ case. Also the maximum of the curve of the normalised EE 
is in the middle of the spin  chain until we have five magnons, but as we increase their number 
this maximum is not in the middle any more. This is the inverse picture with respect to the observations 
of the $SU(2)$ sector, for the normalised EE.

The EE of the of the $SU(1|1)$ sector has some differences with respect to the other two, namely of the 
$SU(2)$ \& $SL(2)$. This is reflected in the right plot of figure \ref{fig:3.4}, where for more than three magnons 
(the case with two magnons and its particularity has been analysed explicitly at the end of section \ref{EE-2magnons})
the pattern we observe is different from the other two sectors.
Here the maximum of the normalised EE is when we cut the shortest possible piece of the spin chain, 
while the minimum is always when we cut the spin chain in the middle. 
In agreement with the observation in the other two sectors, increasing the number of magnons decreases the normalised EE. 

Closing this section we should mention that
the two features that remain universal, that is independent of the particular excited state we consider, are the following:
\begin{itemize}
\item The fact that the EE per magnon decreases as we increase the number of the magnons\footnote{Notice that
for every value of the normalised splitting $N/L$ the behaviour of the normalised entropy and the EE per magnon is the same.}.
\item For the bosonic subsectors $SU(2)$ \& $SL(2)$ the behaviour of the EE presented in figures \ref{fig:3.1}, \ref{fig:3.2} 
and \ref{fig:3.3} is universal, as long as one remains within the broad class of two cut solutions. 
These solutions are of special interest for the thermodynamic limit and the AdS/CFT correspondence.
\end{itemize}


\section{Conclusions and future directions}
\label{Conclusions}

The aim of this paper was to exploit integrability in order to shed some light to the behaviour of the 
Entanglement and Renyi Entropies of the ${\cal N}=4$ SYM spin chain.
Generically, the Entanglement and Renyi Entropies depend on the lengths of the spin 
chain and of the domain of it that we cut, as well as on the details of the excited state whose entropy we are after. 
These details are the number of propagating magnons, their rapidities and the particular sector on which we focus.
Furthermore, since the dilatation operator of  ${\cal N}=4$ receives quantum corrections its eigenstates and the associated 
with them Entanglement and Renyi Entropies will depend on the 't Hooft coupling $\lambda$.  
Our goal was to address these questions about how the EE depends on the aforementioned parameters.   

After providing the reader with the results for the EE of the vacuum and the single magnon state we analytically 
calculated the EE of excited states with two magnons in all closed rank one subsectors of  ${\cal N}=4$ SYM, 
namely $SU(2)$, $SU(1|1)$ and $SL(2)$. 
Our calculation was performed using the formalism of the Coordinate Bethe Ansatz and was leading in the 
coupling expansion (our states were eigenstates of the one-loop dilatation operators) 
but exact in the length of the spin chain and of the part of it we cut. 

In Section \ref{EE-BMN} we calculated the EE of the superconformal primary operator with two excitations in the BMN limit. 
We derived an analytic expression for the EE which is exact in the coupling 
$\lambda'=\frac{g_{YM}^2N}{J^2}=\frac{\lambda}{J^2}$ and interpolates between the weak $\lambda'=0$ and strong
coupling regimes $\lambda'\rightarrow \infty$. This allowed us to analyse the effect of long-range interactions 
of the spin chain on the EE. We have found that the EE of a part of the spin chain is a monotonically increasing 
function of the coupling which saturates to a constant value as 
$\lambda' \rightarrow \infty$ when we keep the length of the chain we cut fixed. This results to a violation of a certain bound for the EE that is present at weak coupling. 
Thus, one of our main conclusions is that, as it is physically anticipated, the entanglement between parts of the 
chain becomes stronger as one increases the coupling $\lambda'$, at least for the superconformal primary operator 
with two excitations.  

In Section \ref{ABA-gen} we employed integrability, and more precisely the powerful formalism of the 
Algebraic Bethe Ansatz in order to calculate numerically the EE of excited states with up to seven 
magnons in the $SU(2)$, $SU(1|1)$ and $SL(2)$ subsectors. 
In the $SU(2)$ and $SL(2)$ subsectors we have focused on excited state corresponding to 
2-cut solutions of the Bethe equations.
Although the absolute value of the EE increases with the number of magnons, its normalised value, that is the EE 
for an $M$ magnon excited state divided by $M$ times the EE of one magnon, decreases with the the number of magnons. 
Some differences in the behaviour of the EE as a function of the magnon number for the two bosonic sectors are scrutinised in Section \ref{ABA-gen}.
The different statistics of the excited states in the $SU(1|1)$ sector lead to a different 
qualitative behaviour of the normalised EE. 
This is the only sector that the normalised EE explicitly saturates the bound for the case of two magnons.  

A number of very interesting questions remain to be answered.
First of all from the perspective of the AdS/CFT correspondence it is important if the calculation of Section \ref{ABA-gen} 
for the $SU(2)$ and $SL(2)$ sectors could be performed for a larger number of magnons in longer spin chains. 
This would allow one to approach the thermodynamic limit in which case the spin chain states will be dual to 
certain semi-classical string solutions. 
However, it seems that for this to be achieved a reformulation of the problem will be needed. In particular one should need to 
obtain expressions for the product of two off-shell states in the thermodynamic limit.
One can then address the question of what is the precise relation between the entropy calculated from the spin chain 
approach and the one which one may calculate from the dual solutions of the non-linear $\sigma$--model
\footnote{One comment is in order. Notice that while in our calculations the EE is always finite (even at the 
BMN limit where $J\rightarrow\infty$) while in all the field theory calculations that appear in the literature 
(see e.g. \cite{Casini:2009sr}) for free field theories the EE is divergent when the cut-off $a$ is sent to zero.}.  
The same question can be asked about the exact in $\lambda'$ EE which we have calculated in Section \ref{EE-BMN}.
Furthermore, it would be interesting to generalise the calculation of Section \ref{EE-BMN} to the case of 
superconformal primary operators with more than two excitations. Another direction would be to calculate the $\frac{1}{J}$ corrections to the BMN EE by considering the near BMN limit.

One could also employ the Perturbative Bethe Ansatz technique to calculate the $g_{YM}^2 N$ corrections to the EE, 
as an order by order expansion in perturbation theory. One should of course interpret the fact that eigenstates of the dilatation 
operators (and as a consequence the corresponding EE) will be scheme-dependent. 
Notice that such a complication is absent both in the BMN limit considered in  Section \ref{EE-BMN}
and in the leading in $g_{YM}^2 N$ calculations of Section \ref{ABA-gen}.


\section*{Acknowledgements}

GG would like to thank K. Sfetsos for enlightening discussions.
DZ acknowledges financial support from the Universitat Internacional de Catalunya.

\appendix

\section{Algebraic Bethe Ansatz: Essential formulas}
\label{app:ABA}

In this appendix we collect all the essential mathematical formulas needed for the construction of the ABA, following 
the references \cite{Escobedo:2010xs,Caetano:2011eb,Escobedo:2012ama}. We introduce the following functions
that determine the rank one subsector of the ${\cal N}=4$ SYM we are working.


\subsection*{$SU(2)$ sector}

The expressions for the basic building blocks are 
\begin{equation} \label{SU2def}
f(u) \equiv 1 \, + \, \frac{i}{u} \, , \quad 
g(u) \equiv \frac{i}{u} \, , \quad 
h(u) \equiv \frac{f(u)}{g(u)}  = \frac{u+i}{i} \, , \quad 
t(u) \equiv \frac{g(u)^2}{f(u)} =  \frac{-1}{u(u+i)} \, ,
\end{equation}
\begin{equation}
a(u) \, \equiv \, \left(u \, + \, \frac{i}{2}\right)^L \, , \quad 
d(u) \, \equiv \, \left(u \, - \, \frac{i}{2}\right)^L,
\end{equation}
while the $SU(2)$ Bethe equations that determine the set of rapidities, are
\begin{equation}
1 \, \equiv  \, \left(\frac{u_j \, + \, i/2}{u_j \, - \, i/2}\right)^L  \prod_{k\neq j}^M \frac{u_j \, - \, u_k\, - \, i}{u_j\, - \, u_k\, +\, i} \,.
\end{equation}


\subsection*{$SL(2)$ sector}

The expressions for the basic building blocks are 
\begin{equation} \label{SL2def}
f(u) \, = \,  1 \, + \, \frac{i}{u} \, , \quad 
g(u) \, = \,  \frac{i}{u} \, , \quad 
a(u) \, = \, \left(u \, - \, \frac{i}{2}\right)^L \, , \quad 
d(u) \, = \, \left(u \, + \, \frac{i}{2}\right)^L,
\end{equation}
while the $SL(2)$ Bethe equations are
\begin{equation}
1 \, \equiv  \, \left(\frac{u_j \, + \, i/2}{u_j \, - \, i/2}\right)^L  \prod_{k\neq j}^M \frac{u_j \, - \, u_k\, + \, i}{u_j\, - \, u_k\, -\, i} \,.
\end{equation}


\subsection*{$SU(1|1)$ sector}

The expressions for the basic building blocks are 
\begin{equation} \label{SU11def}
f(u) \, = \, \frac{i}{u} \, , \quad 
g(u) \, = \,  \frac{i}{u} \, , \quad 
a(u) \, = \, \left(u \, + \, \frac{i}{2}\right)^L \, , \quad 
d(u) \, = \, \left(u \, - \, \frac{i}{2}\right)^L,
\end{equation}
while the $SU(1|1)$ Bethe equations are
\begin{equation}
1 \, \equiv  \, \left(\frac{u_j \, + \, i/2}{u_j \, - \, i/2}\right)^L  \, .
\end{equation}
%
%
In order to simplify the expressions we introduce the following shorthand notation
\begin{equation} \label{notation}
F^a \, \equiv \, \prod_{u_j \in \, a} F(u_j) \, ,\qquad 
F^{a\bar a} \, \equiv \!\!\! 
\prod_{\scriptsize \begin{array}{c} {u_i} \in a \\ v_j \in \bar a \end{array}} \!\!\!F(u_i-v_j)\,  ,\qquad 
F_<^{aa} \, \equiv \!\!\! 
\prod_{\scriptsize \begin{array}{c} {u_i, u_j} \in a  \\  i<j \end{array}} \!\!\!F(u_i-u_j) \, .
\end{equation}
Now we can write the expression for the function $H(a,\bar{a})$, that weights the different partitions of the 
spin chain, according to the algebraic normalizations of \cite{Escobedo:2010xs}
\begin{equation} \label{Hdef}
H(a,\bar{a}) \, = \, f^{a \bar{a}} \, d_r^a \, a_l^{\bar{a}}\, , 
\end{equation}
where $a_{l}$ and $d_r$ are defined as in equations \eqref{SU2def}, \eqref{SL2def} \& \eqref{SU11def}, 
but using instead of $L$ the lengths for the left and the right subchain respectively.


\subsection*{Scalar Product}

In order to compute the RDM we need to evaluate the scalar product between two Bethe 
wavefunctions for arbitrary $\{u\}$ and $\{v\}$. A recursion relation for such an expression is computed analytically 
in references \cite{Escobedo:2010xs} and \cite{Caetano:2011eb}.  
Here, for completeness, we present the outcome of that computation.

Consider two Bethe states, which are parametrized by $u_i$ and $v_i$, with $i=1,\dots,N$. 
The scalar product $S_N (\{v\},\{u\}) \equiv \< \{v^*\} | \{u\} \>$ is given by the following recursion relation
\begin{eqnarray} \label{rec_relation}
&&S_N \left(\{v_1,\dots,v_N\},\{u_1,\dots,u_N\}\right) = \sum_{n} b_n \, S_{N-1} 
\left(\{v_1, \dots, \hat v_n,\dots,v_N\},\{\hat u_1,u_2, \dots,u_N\}\right)
\nonumber  \\[2pt]
&&\qquad \qquad - \sum_ {n<m}  c_{n,m} \, S_{N-1}
\left(\{ u_1,v_1,\dots \hat v_{n},  \dots , \hat v_{m},\dots v_N \}, \{\hat u_1,u_2, \dots,u_N\}\right) \, , 
\end{eqnarray}
where a Bethe root with a hat means that it is omitted. The definitions for $b_n$ and $c_{n,m}$ are
the following
\begin{equation}
b_n \, = \, g(u_1-v_n) a(v_n) d(u_1) \prod_{j\neq n}^N \, f(u_1-v_j) \, f(v_j-v_n)  \, + \,  \left(u_1 \leftrightarrow v_n\right) \, , 
\end{equation}
\begin{equation}
c_{n,m} \, = \, {g(u_1-v_{n}) \, g(u_1-v_{m}) \, a(v_{m}) \, d(v_{n})} \, {f(v_{n}-v_{m})} 
\prod_{j\neq n,m}^N f(v_{n} - v_j)  \, f(v_j - v_{m}) \, + \, \left(n \leftrightarrow m\right).
\end{equation}
Using \eqref{rec_relation} and substituting the corresponding expressions for the functions $f(u),g(u),a(u)$ and $d(u)$ 
(from either \eqref{SU2def}, \eqref{SL2def} or \eqref{SU11def}), it is possible to calculate the scalar product 
for any of the rank one subsectrors. 


\section{Bethe Roots}
\label{app:BR}

In section \ref{ABA-gen} we calculate the EE of excited states that belong either on the $SU(2)$ or on the $SL(2)$ or 
on the $SU(1|1)$ sector. In this appendix we list the Bethe roots of these excited states and in the $SU(2)$ case we also
plot them, since they distribute themselves along two disjoint cuts on the complex plane. 

We start from the Bethe roots of the excited states of figures \ref{fig:3.1} and \ref{fig:3.2}. 
In those figures we are considering spin chains with 12, 14, 16, 18 \& 20 sites 
with a number of magnons ranging from $2$ to $7$. The Bethe roots for these 
magnons are the following
\\ \\
{\small
\underline{$SU(2)$ sector}: {\bf L=12} \\[3pt]
2 m $\Rightarrow$ \verb" {-1.703, 1.703}",\\[3pt]
3 m $\Rightarrow$ \verb" {-0.478 + 0.500 I, -0.478 - 0.500 I, 1.418}" \\[3pt]
4 m $\Rightarrow$ \verb" {-1.296 + 0.564 I, -1.296 - 0.564 I, 1.296 - 0.564 I, 1.296 + 0.564 I}" \\[3pt]
5 m $\Rightarrow$ \verb" {-1.11 + 0.537 I, -1.11 - 0.537 I, 0.979 - I, 1.009, 0.979 + I}" \\[3pt]
6 m $\Rightarrow$ \verb" {-0.676 + I, -0.676, -0.676 - I, 0.676 - I, 0.676, 0.676 + I}"
}
\\ \\
{\small
\underline{$SU(2)$ sector}: {\bf L=14} \\[3pt]
2 m $\Rightarrow$ \verb" {-2.029, 2.029}",\\[3pt]
3 m $\Rightarrow$ \verb" {-0.713 + 0.501 I, -0.713 - 0.501 I, 1.754}" \\[3pt]
4 m $\Rightarrow$ \verb" {-1.644 + 0.601 I, -1.644 - 0.601 I, 1.644 - 0.601 I, 1.644 + 0.601 I}" \\[3pt]
5 m $\Rightarrow$ \verb" {-1.47 + 0.574 I, -1.47 - 0.574 I, 1.428, 1.37 - 1.043 I, 1.37 + 1.043 I}" \\[3pt]
6 m $\Rightarrow$ \verb" {-1.16 + 1.01 I, -1.2, -1.16 - 1.01 I, 1.16 - 1.01 I, 1.2, 1.16 + 1.01 I}"\\[3pt]
7 m $\Rightarrow$ \verb" {-0.898 + 0.994 I, -0.898 - 0.994 I, 0.686 + 1.499 I, 0.686 - 1.499 I, "\\[3pt]
    \verb"         0.665 - 0.5 I, 0.665 + 0.5 I, -0.907}"
}
\\ \\
{\small
\underline{$SU(2)$ sector}: {\bf L=16} \\[3pt]
2 m $\Rightarrow$ \verb" {-2.352, 2.352}",\\[3pt]
3 m $\Rightarrow$ \verb" {-0.922 + 0.502 I, -0.922 - 0.502 I, 2.083}" \\[3pt]
4 m $\Rightarrow$ \verb" {-1.981 + 0.639 I, -1.981 - 0.639 I, 1.981 - 0.639 I, 1.981 + 0.639 I}" \\[3pt]
5 m $\Rightarrow$ \verb" {-1.815 + 0.612 I, -1.815 - 0.612 I, 1.721 - 1.1 I, 1.804, 1.721 + 1.1 I}" \\[3pt]
6 m $\Rightarrow$ \verb" {-1.54 + 1.06 I, -1.6, -1.54 - 1.06 I, 1.54 - 1.06 I, 1.6, 1.54 + 1.06 I}"\\[3pt]
7 m $\Rightarrow$ \verb" {-1.351 + 1.023 I, -1.396, -1.351 - 1.023 I, 1.230 - 1.464 I, "\\[3pt]
    \verb"        1.262 - 0.502 I, 1.262 + 0.502 I, 1.23 + 1.464 I}"
}
\\ \\ 
{\small
\underline{$SU(2)$ sector}: {\bf L=18} \\[3pt]
2 m $\Rightarrow$ \verb" {-2.675, 2.675}",\\[3pt]
3 m $\Rightarrow$ \verb" {-1.117 + 0.505 I, -1.117 - 0.505 I, 2.408}" \\[3pt]
4 m $\Rightarrow$ \verb" {-2.312 + 0.675 I, -2.312 - 0.675 I, 2.312 - 0.675 I, 2.312 + 0.675 I}" \\[3pt]
5 m $\Rightarrow$ \verb" {-2.15 + 0.65 I, -2.15 - 0.65 I, 2.063 - 1.156 I, 2.159, 2.063 + 1.156 I}" \\[3pt]
6 m $\Rightarrow$ \verb" {-1.89 + 1.1 I, -1.98, -1.89 - 1.1 I, 1.89 - 1.1 I, 1.98, 1.89 + 1.1 I}"\\[3pt]
7 m $\Rightarrow$ \verb" {1.614 - 1.503 I, 1.702 - 0.511 I, 1.702 + 0.511 I, 1.614 + 1.503 I, "\\[3pt]
    \verb"        -1.719 + 1.074 I, -1.790, -1.719 - 1.074 I}"
}
\\ \\ 
{\small
\underline{$SU(2)$ sector}: {\bf L=20} \\[3pt]
2 m $\Rightarrow$ \verb" {-2.996, 2.996}",\\[3pt]
3 m $\Rightarrow$ \verb" {-1.302 + 0.51 I, -1.302 - 0.51 I, 2.732}" \\[3pt]
4 m $\Rightarrow$ \verb" {-2.64 + 0.71 I, -2.64 - 0.71 I, 2.64 - 0.71 I, 2.64 + 0.71 I}" \\[3pt]
5 m $\Rightarrow$ \verb" {-2.48 + 0.685 I, -2.48 - 0.685 I, 2.397 - 1.214 I, 2.5, 2.397 + 1.214 I}" \\[3pt]
6 m $\Rightarrow$ \verb" {-2.23 + 1.17 I, -2.3, -2.23 - 1.17 I, 2.23 - 1.17 I, 2.3, 2.23 + 1.17 I}"\\[3pt]
7 m $\Rightarrow$ \verb" {1.965 - 1.567 I, 2.090 - 0.525 I, 2.090 + 0.525 I, 1.965 + 1.567 I, "\\[3pt]
    \verb"        -2.066 + 1.131 I, -2.153, -2.066 - 1.131 I}"
}
\\ \\ 
We only plot the Bethe roots that correspond to a spin chain with 14 sites, since all the other plots are similar.
\begin{figure}[h] 
   \centering
      \includegraphics[width=11cm]{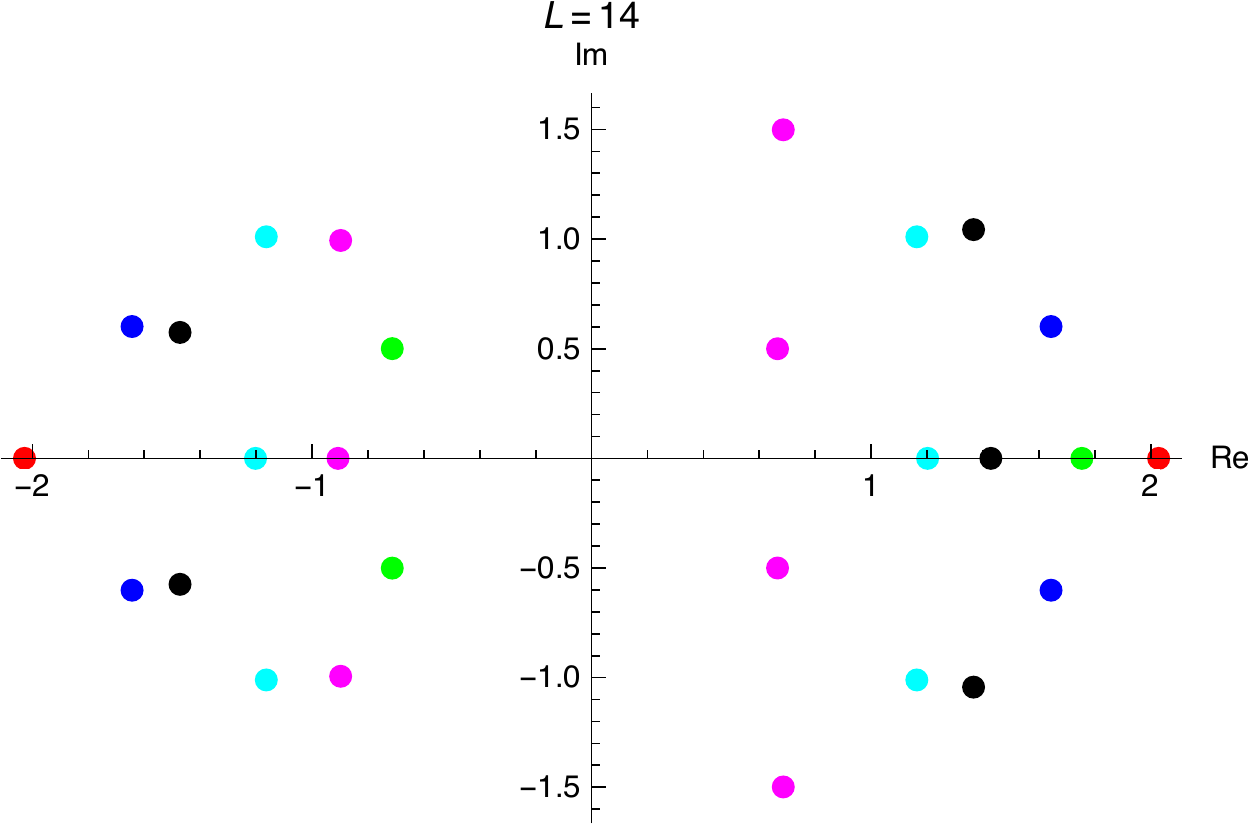}\\
       \caption{The correspondence between colour and number of magnons is the following: Red $\Rightarrow$ 2 magnons, 
       Green $\Rightarrow$ 3 magnons, Blue $\Rightarrow$ 4 magnons, Black $\Rightarrow$ 5 magnons, 
       Cyan $\Rightarrow$ 6 magnons \& Magenta $\Rightarrow$ 7 magnons.}
   \label{fig:app.1}
\end{figure}

The Bethe roots that correspond to the excited states of the figures \ref{fig:3.3} ($SL(2)$ sector) and 
\ref{fig:3.4} ($SU(1|1)$ sector), for a spin chain with with $14$ sites, are the following 
\\ \\
{\small
\underline{$SL(2)$ sector}\\[3pt]
2 m $\Rightarrow$ \verb" {-2.352, 2.352}",\\[3pt]
3 m $\Rightarrow$ \verb" {-1.889, -3.069, 2.521}" \\[3pt]
4 m $\Rightarrow$ \verb" {-2.024, -3.268, 3.268, 2.024}" \\[3pt]
5 m $\Rightarrow$ \verb" {-2.166, -3.472, 3.946, 2.652, 1.726}"\\[3pt]
6 m $\Rightarrow$ \verb" {-1.845, 1.845, -2.822, -4.17, 4.17, 2.822}"\\[3pt]
7 m $\Rightarrow$ \verb" {4.831, 1.619, 2.415, 3.437, -1.97, -2.998, -4.398}" \\
}\\
and
\\ \\
{\small
\underline{$SU(1|1)$ sector}\\[3pt]
3 m $\Rightarrow$ \verb" {0.399, 0.627, 1.038}" \\[3pt]
4 m $\Rightarrow$ \verb" {0, 0.114, 0.241, 0.399, 0.627}" \\[3pt]
5 m $\Rightarrow$ \verb" {-2.166, -3.472, 3.946, 2.652, 1.726}"\\[3pt]
6 m $\Rightarrow$ \verb" {-0.175, -0.056, 0.056, 0.175, 0.314, 0.5}"\\[3pt]
7 m $\Rightarrow$ \verb" {-0.399, -0.241, -0.114, 0, 0.114, 0.241, 0.399}" \\
}\\


\end{document}